\documentclass[a4paper,fleqn,usenatbib]{mnras}

\usepackage[T1]{fontenc}
\usepackage{ae,aecompl}

\usepackage{newtxtext,newtxmath}

\usepackage{graphicx}	
\usepackage{amsmath}	
\usepackage{amssymb}	
\usepackage{siunitx, tensor, subfig, footmisc, tabu, float, flafter}
\usepackage[inline]{enumitem}
\usepackage[capitalise,sort]{cleveref}

\newcolumntype{'}{!{\vrule width 2pt}}

\newcommand{\Dif}{\mathrm{d}}

\DeclareMathOperator{\Msun}{\si{M_\odot}}
\DeclareMathOperator{\Lsun}{\si{L_\odot}}
{}
{}
{}
{}
{}

\newcommand{\tfo}[1]{\texttt{#1}}
\newcommand{\tso}[1]{\textsc{#1}}

\newcommand{\shw}{Schwarzschild}
\newcommand{\SZ}{\({\rm S}0\)}
\newcommand{\mgeS}{MGE\(_\mu\)}
\newcommand{\mgeT}{MGE\(_\Sigma\)}
\newcommand{\atomic}[2]{\nuclide{#1}\textsc{#2}}

\usepackage{etoolbox}
\makeatletter
\patchcmd{\hyper@makecurrent}{%
    \ifx\Hy@param\Hy@chapterstring
        \let\Hy@param\Hy@chapapp
    \fi
}{%
    \iftoggle{inappendix}{
        \@checkappendixparam{chapter}%
        \@checkappendixparam{section}%
        \@checkappendixparam{subsection}%
        \@checkappendixparam{subsubsection}%
        \@checkappendixparam{paragraph}%
        \@checkappendixparam{subparagraph}%
    }{}%
}{}{\errmessage{failed to patch}}

\newcommand*{\@checkappendixparam}[1]{%
    \def\@checkappendixparamtmp{#1}%
    \ifx\Hy@param\@checkappendixparamtmp
        \let\Hy@param\Hy@appendixstring
    \fi
}
\makeatletter

\newtoggle{inappendix}
\togglefalse{inappendix}

\apptocmd{\appendix}{\toggletrue{inappendix}}{}{\errmessage{failed to patch}}

\LetLtxMacro{\oldsqrt}{\sqrt} 

\title[NGC 3115]{Combining Stellar Populations with Orbit-Superposition Dynamical Modelling - the Formation History of the Lenticular Galaxy NGC 3115}

\author[A. Poci et al.]{
Adriano Poci$^{1}$\thanks{E-mail: adriano.poci@students.mq.edu.au (MQU)},
Richard M. McDermid$^{1}$,
Ling Zhu$^{2}$,
Glenn van de Ven$^{3,4}$
\\
$^{1}$Astronomy, Astrophysics, and Astrophotonics Research Centre, Department of Physics and Astronomy, Macquarie University, NSW 2109, Australia\\
$^{2}$Shanghai Astronomical Observatory, Chinese Academy of Sciences, 80 Nandan Road, Shanghai 200030, China \\
$^{3}$Department of Astrophysics, University of Vienna, T\"urkenschanzstrasse 17, 1180 Vienna, Austria\\
$^{4}$European Southern Observatory, Karl-Schwarzschild-Str 2, D-85748 Garching bei Munchen, Germany
}

\date{Accepted XXX. Received YYY; in original form ZZZ}

\pubyear{2019}

\begin{document}
\label{firstpage}
\pagerange{\pageref{firstpage}--\pageref{lastpage}}
\maketitle

\begin{abstract}
We present a combination of the \shw\ orbit-superposition dynamical modelling technique with the spatially-resolved mean stellar age and metallicity maps to uncover the formation history of galaxies. We apply this new approach to a remarkable 5-pointing mosaic of VLT/MUSE observations obtained by \cite{guerou16} extending to a maximum galactocentric distance of \(\sim \SI{120}{\arcsecond} \left(\SI{5.6}{kpc}\right)\) along the major axis, corresponding to \(\sim 2.5~\si{R_e}\). Our method first identifies `families' of orbits from the dynamical model that represent dynamically-distinct structures of the galaxy. Individual ages and metallicities of these components are then fit for using the stellar-population information. Our results highlight components of the galaxy that are distinct in the combined stellar dynamics/populations space, which implies distinct formation paths. We find evidence for a dynamically-cold, metal-rich disk, consistent with a gradual in-situ formation. This disk is embedded in a generally-old population of stars, with kinematics ranging from dispersion-dominated in the centre to an old, diffuse, metal-poor stellar halo at the extremities. We find also a direct correlation between the dominant dynamical support of these components, and their associated age, akin to the relation observed in the Milky Way. This approach not only provides a powerful model for inferring the formation history of external galaxies, but also paves the way to a complete population-dynamical model.
\end{abstract}




\section{Introduction}\label{sec:intr}
The present-day observed state of a galaxy is the result of the integration over its entire formation history, including external accretion/mergers, in-situ star-formation, and passive stellar evolution. To determine how and when a galaxy has built up its mass, it is necessary to disentangle its present-day state into spatially- and chemically-distinct events. Typically, the studies of stellar populations and dynamical properties have remained independent, however it is the union of these two aspects that is necessary to be able to investigate the origin of a galaxy's mass over its formation history. These ideas have been investigated for some time for the Milky Way, where a wealth of chemical and dynamical information can be directly obtained for individual stars. We endeavour here to extend these analyses to external, unresolved galaxies.\par
Historically, dynamical models of galaxies have been utilised for a wide range of applications, from individual galaxy analyses to large statistical samples of the galaxy population. These efforts have attempted to place constraints on, among other properties, the mass of the super-massive black holes (SMBH) at the centres of external galaxies (see \citealp{kormendy13} and references in Table 1 therein, and more recently \citealp{seth14,krajnovic18}); the IMF mass normalisation \citep[for instance, see][]{thomas11, cappellari12, lyubenova16, davis17, li17, oldham18a}; the dark-matter content/distribution in galaxies using stellar kinematics \citep[for example, see][]{cappellari13a, tortora19} all the way down in mass to dwarf spheroidals \citep{jardel12}, and using gas kinematics \citep[for example, see][]{corbelli00,gentile04,diteodoro19}. Dynamical models have also been used to uncover new relationships between galaxy structural parameters, including the widely-used stellar spin-ellipticity \((\lambda_r-\epsilon)\) correlation \citep[introduced in][]{emsellem07}, as well as other observed correlations with dynamical properties \citep{cappellari16}. Finally, dynamical models of individual galaxies have been used to probe internal dynamical structures in great detail \citep[for instance, see][]{krajnovic05, vandenbosch08, krajnovic15, zhu16b}, placing strong constraints on the formation mechanisms of those specific galaxies.\par
Independently, stellar-population models of galaxies have been utilised for their own range of applications, covering similar scales in sample size. These models have been applied to constrain the IMF shape, normalisation, and low-mass cut-off \citep[for instance, see][]{conroy12, smith12, martinnavarro15, alton17, rosani18, dries18, vaughan18}; the measurement and interpretation of \(\alpha\)-enhancement \citep[as in][]{thomas05,conroy14, greene15, mcdermid15}; the measurement of spatial gradients in stellar properties \citep[for instance, see][]{mehlert03, sanchezblazquez07, kuntschner10, li18}; assembly timescales for galaxy formation \citep{martinnavarro18}. Similarly to dynamical models, stellar population models have been used to uncover new relationships between galaxy structural parameters through a variety of scaling relations \citep[see][for some such relations]{gallazzi08, graves09, graves09b, mcdermid15, walcher15}. Again, stellar population models of individual (or handfuls of) galaxies in great detail offer insight into the specific formation path of these galaxies, as well as the presence of sub-structures with distinct stellar populations \citep[for instance, see][]{barbosa16, mentz16, streich16}.\par
Driven by the influx of spatially-resolved observations coming from integral-field units (IFU), more recent investigations have focused on attempting to infer structural, dynamical and/or chemical properties for localised regions of galaxies, by decomposing them into physically-motivated components. In fact, this concept pre-dates the large-scale use of IFU with works dealing with, for instance, photometric disk/bulge decompositions of surface brightness profiles \citep[for example, see][]{kent85, cinzano93, scorza95, moriondo98, krajnovic13}, including using multiple filters \citep{dimauro18}. Aside from surface brightness, radial profiles of other parameters have also been subject to analogous decompositions. For instance, the decomposition of gas (usually \(\atomic{H}{i}\)) circular velocity profiles into contributions from different galaxy sub-components is a well-established practise \citep[as in][]{vanalbada85, carignan88, battaglia06, noordereer07, swaters12, sofue17, aniyan16, aniyan18}, while decompositions of mass profiles have attempted to infer the contributions from dark matter and baryons \citep[stars and globular clusters, gas, et cetera;][]{poci17, annunziatella17, bellstedt18}. These concepts have been extended to two dimensions, including multi-band photometric disk/bulge decompositions of images, rather than profiles \citep[for instance, see][]{scorza98,desouza04, norris06, simard11, mendezabreu17, dallabonta18}. Moreover, there have been recent efforts to conduct the decomposition directly on an observed spectrum \citep{johnston12, coccato15, tabor17, coccato18}, to similarly determine the contributions to various spectral features coming from `distinct' galaxy subcomponents. This type of component-based approach attempts to isolate the distinct contributions to observed galaxies from regions which may or may not have had different origins and/or formation paths, however they have thus far dealt with the problem from only one perspective - dynamics \emph{or} stellar populations.\par
These works, and others, have motivated a need for combining the aforementioned dynamical and chemical models in order to investigate the full formation history of a galaxy. This combination is in fact necessary to be able to determine the origin of the different components of a galaxy. There have been only a small number of works in which different chemical and dynamical populations are simultaneously treated, such as the models of \cite{zhu16, zhu16b}, which use globular clusters (GC) as discrete tracers of the kinematics, while simultaneously fitting for two (chemical) populations of GC \citep[based on the discrete modelling prescription of][]{watkins13}. This showed that there are observable distinctions to be made in the combined populations/dynamics space (and indeed physical space) between different components of galaxies.\par
In this paper, we describe a new approach which aims to decompose a galaxy into dynamical and chemical components in order to infer its formation history. We present also an application of this method to a real galaxy; the nearby lenticular galaxy NGC 3115. \cref*{sec:data} describes the observational data used in this work. \cref*{sec:srsfh,sec:dyn} describe in detail the multiple aspects of our chemical and dynamical analyses, respectively. The results of our show case are presented in \cref*{sec:results}. \cref*{sec:formation} presents the formation history of NGC 3115 as determined from our method, and the connections between dynamical and chemical properties.\par

\section{Observational Data and Kinematic Extraction}\label{sec:data}
The observations used in this work were obtained and reduced by \cite{guerou16}. NGC 3115 is the nearest lenticular (\SZ) galaxy to the Milky Way, with an orientation very close to edge-on. \cite{tonry01} used surface brightness fluctuations to measure a distance modulus for NGC 3115 of \((m-M)=29.93\pm 0.09\), placing it at a physical distance of \(9.7~\si{Mpc}\). It has an effective (half-light) radius of \(R_e = 47.32\si{\arcsecond}\big/2.23~\si{kpc}\) \citep{emsellem99}. Our data extends to a galactocentric radius of \(\sim \SI{120}{\arcsecond} \left(\SI{5.6}{kpc}\right)\) along the major axis, and so we have coverage out to \(\sim 2.5 R_e\).\par
The data set has a pixel-scale of \(\SI{0.2}{\arcsecond\ pixel^{-1}}\ (\SI{9.4}{pc\ pixel^{-1}})\), and consists of over \(360,000\) individual spectra. We refer the interested reader to \cite{guerou16} for further details regarding the observational procedure and data-reduction techniques.\par
For all subsequent analysis, we consider only the spectral range \(\lambda \in \left[4760, 6400\right]~\si{\angstrom}\). This is to reduce the impact of residual sky emission lines on the extracted stellar kinematic and population properties. We spatially bin the reduced datacube using a Python implementation\footnote{\label{fn:code}Available at \href{http://www-astro.physics.ox.ac.uk/~mxc/software/}{http://www-astro.physics.ox.ac.uk/\textasciitilde mxc/software/}} of the Voronoi tessellation algorithm \citep{cappellari03} to a target signal-to-noise ratio \((S/N)\) of 90 per bin. This relatively high threshold is set to ensure an accurate recovery of (the moments of) the line-of-sight velocity distribution (LOSVD), as well as the subsequent stellar population analyses.\par
Although kinematics were extracted by \cite{guerou16}, we re-derive the kinematics\footnote{\label{fn:exData}These data products are also available at \href{https://datacentral.org.au/teamdata/NGC3115/public/}{https://datacentral.org.au/teamdata/NGC3115/public/}} here with the new \(S/N\) threshold for the spatial binning. This allows us to extract the first six Gauss-Hermite coefficients of the parametrised LOSVD, which provide important additional constraints on the dynamical model.\par
For the kinematic extraction, we use a Python implementation\footref{fn:code} of the parametric Penalised PiXel-Fitting code {\sc ppxf} \citep{cappellari04, cappellari16}, with 985 stellar templates from the empirical MILES library \citep{sanchezblazquez06, falconbarroso11}. {\sc ppxf} finds the linear combination of the provided templates that, when convolved with an LOSVD parametrised by the 6 Gauss-Hermite coefficients, best matches the observed spectrum. Computing such a fit for every (Voronoi-binned) spectrum in the cube provides the best-fitting LOSVD (or parametrisation thereof) at every spatial location. Since the spectral resolutions of the MILES models and MUSE data are comparable, both sets of spectra are kept at their native resolution. Every bin is fit with the freedom of the full template library, and with an additive polynomial of order 16. This combination of additive polynomial and stellar templates is used to achieve the best possible fit to the spectrum, in order to most accurately recover the LOSVD, without being tied to any particular stellar population model. During the fitting, spurious spectral pixels / artefacts in the data are iteratively clipped. Errors for all 6 Gauss-Hermite moments are computed by Monte Carlo simulations on every spectrum individually for which kinematics are extracted. We derive a mean uncertainty (and apply a floor on these uncertainties during the dynamical model described in \cref*{sec:dyn}), for \(V\) and \(\sigma\) of \(10.30\ (7.00)\) and \(3.17\ (2.00)\ \si{\kilo\metre\per\second}\), respectively, and for moments \(3\) through \(6\) of \(0.0016\ (0.0016)\), \(0.0020\ (0.0019)\), \(0.0016\ (0.0016)\), \(0.0021\ (0.0020)\). The floor on the uncertainties prevents the dynamical model from preferentially fitting the central pixels (which have the smallest errors).

\section{Spatially-Resolved Star-Formation History}\label{sec:srsfh}
Our analysis of NGC 3115 requires the computation of a spatially-resolved star-formation history (SFH). To do this, we employ full-spectral-fitting techniques, with the aim of investigating chemically-distinct components in our dataset. To fit the spectra, we again utilise {\sc ppxf}, but use the MILES-IndoUS-CaT (MIUSCAT) Single Stellar Population (SSP) model library \citep{vazdekis10} as templates. While we aim to compare our results with those from \cite{guerou16}, we adopt different SSP templates for our SFH. From an initial run of the dynamical model, we inferred a dynamical (enclosed) mass that was lower than the stellar mass derived assuming a \cite{salpeter55} IMF \citep[which was the assumption in][]{guerou16}. Such seemingly non-physical discrepancies have been found previously \citep{lyubenova16}, and is interpreted as evidence to exclude single-power-law IMF shapes. We therefore assume a \cite{kroupa02a} IMF for all stellar-population models in this work. The SSP library used here is based on the Padova isochrones \citep{girardi00}, with ages \(0.1 \leq t \leq 14.1\ \si{Gyr}\) and metallicities \(-1.71 \leq Z \leq 0.22\ \si{dex}\)\footnote{Determined from the ``safe ranges'' taken from the \href{http://www.iac.es/proyecto/miles/pages/ssp-models/safe-ranges.php}{MILES} website over the MUSE wavelength range}. These models do not consider any \(\alpha\) abundances explicitly, and since they are based on empirical stars, therefore share the same \(\alpha\)-enhancement characteristics as the solar neighbourhood. During the {\sc ppxf} fit, we employ a small fractional regularisation to reduce the intrinsic degeneracy of the fit. This regularisation prefers a smooth solution that would otherwise be degenerate with a `spiky' SFH. We do not use any additive polynomial, as this would change the relative strengths of the spectral features, which would in turn impact the derived stellar population properties. We do, however, employ a multiplicative polynomial of order 16 in order to account for the continuum. The LOSVD are left free during the stellar-population fits. This is done to minimise any possible systematics and template mismatch, while maintaining good fits across the FOV. We confirmed that the resulting SFH\footref{fn:exData} is consistent within the errors when the kinematics were fixed to those extracted in \cref*{sec:data}.\par
\begin{figure}
    \centerline{
        \includegraphics[width=0.5\columnwidth]{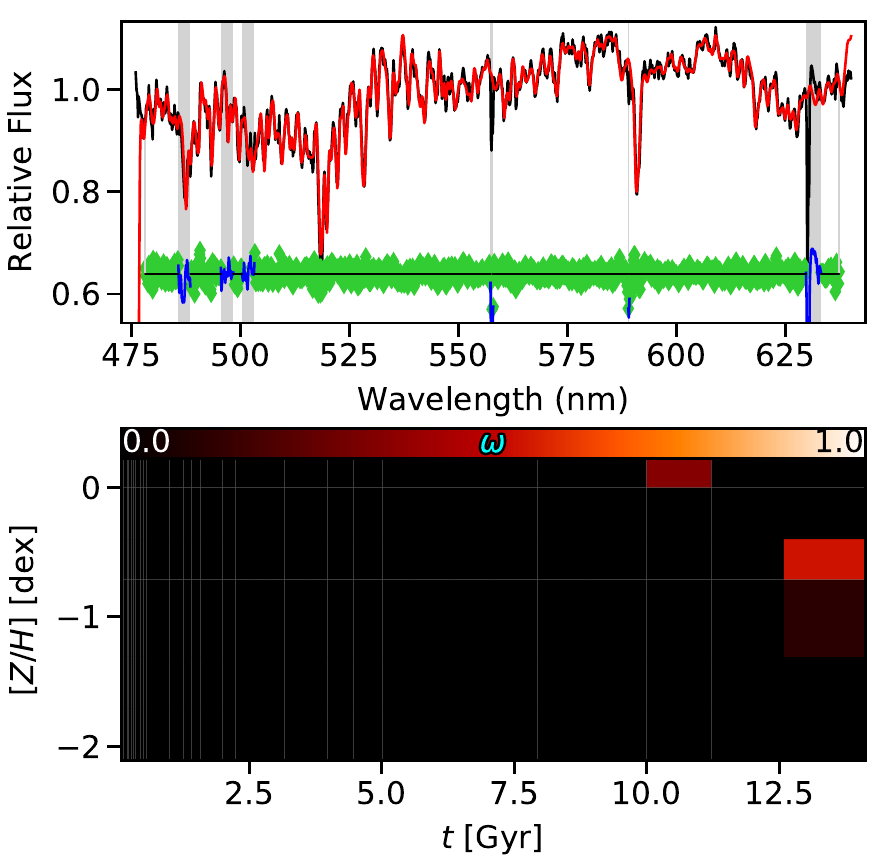}
        \includegraphics[width=0.5\columnwidth]{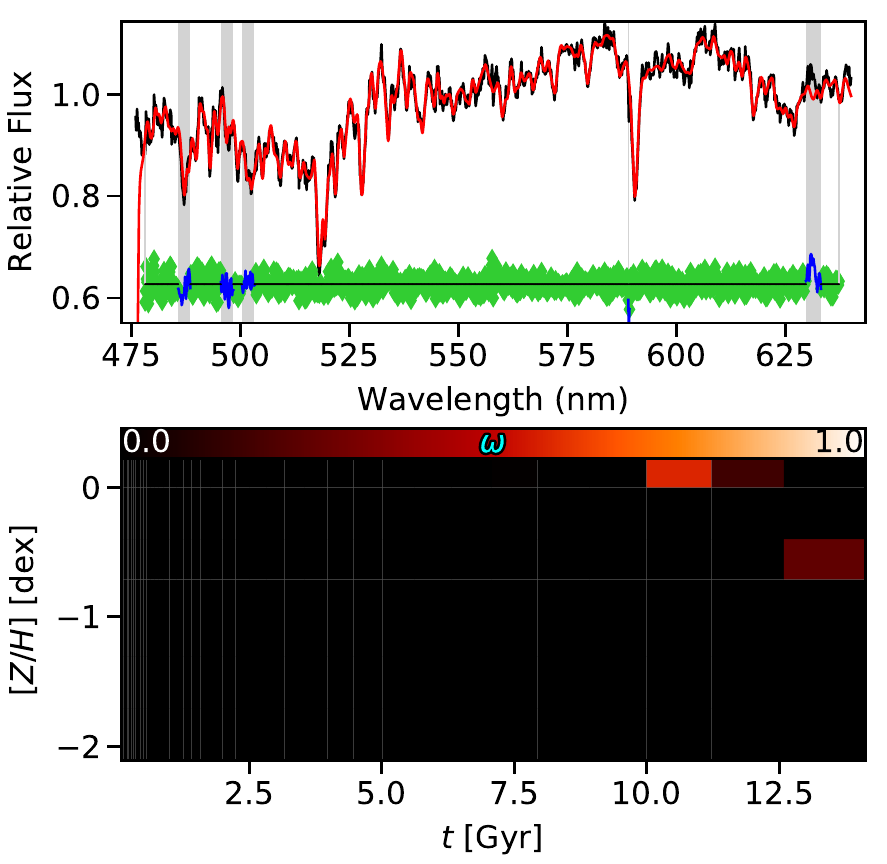}
    }
    \caption{The fit in {\sc ppxf} to the highest (left) and lowest (right) \(S/N\) spectra in the dataset, with \(S/N=119\) and \(67\), respectively. In each panel, the top row shows the data and model in black and red, respectively. The residuals are shown as green diamonds, and have been arbitrarily offset for presentation. The grey bands are masked during the fit. The outer-most pixels on either extremity have also been excluded to avoid edge effects. The bottom panels show the associated luminosity-weighted SFH, which illustrates the contributions from each age/metallicity bin to the corresponding spectrum.}
    \label{img:regSpec}
\end{figure}

\cref*{img:regSpec} illustrates this fitting concept on the highest and lowest \(S/N\) spectra in the dataset, as well as the associated SSP weights for the ages, \(t\), and metallicities, \(Z\).
We compute similar fits for every binned spectrum in the datacube to investigate the spatial behaviour of the stellar populations. For reasons discussed in \cref*{ssec:orbSol}, we compute {\em luminosity}-weighted SSP properties by removing the relative normalisation of all the template spectra prior to fitting.

\section{Dynamical Model}\label{sec:dyn}
In order to identify intrinsically-distinct components, we employ the \shw\ orbit-superposition technique \citep{schwarzschild79}. In this work we use a triaxial extension of the axisymmetric method originally presented in \cite{vandermarel98} and \cite{cretton99b}, and developed further by \cite{cappellari06}. This triaxial implementation is detailed extensively in \cite{vandeven08,vandenbosch08} and developed further by \cite{zhu18, zhu18b}. We present here a brief summary of the relevant aspects of this implementation, and refer the reader to the above references for further details.
\subsection{Mass Model}\label{ssec:mass}
The \shw\ orbit-superposition method generates galaxy models within a given stationary gravitational potential. Fitting \shw\ models to observations therefore requires finding the best input gravitational potential that reproduces the observable constraints. Since the gravitational potential can not be measured directly, we construct each input mass model as the combined contributions from a stellar mass model, a dark matter parametrisation, and a point-source central super-massive black-hole (SMBH) component.
\subsubsection{Stellar-Mass Model}\label{sssec:stellarMassModel}
\begin{figure}
    \centerline{
        \includegraphics[width=1.0\columnwidth]{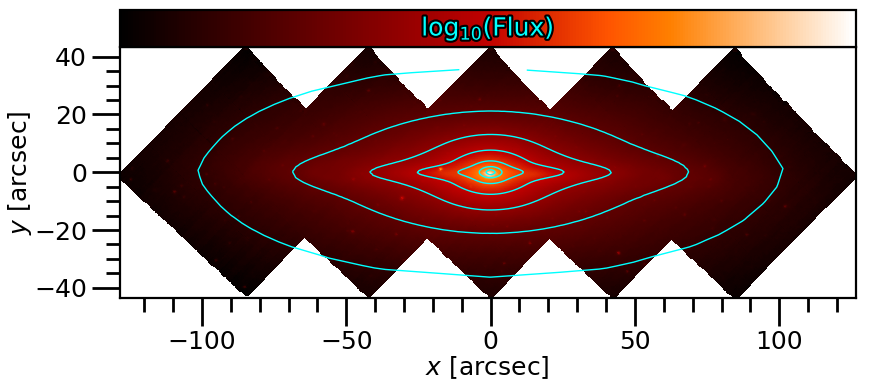}
    }
    \caption{The reconstructed image from the \(3{\rm D}\) data cube in arbitrary flux units, and the surface-brightness MGE contours overplotted in \(1~\si{mag}\) intervals.}
    \label{img:mge}
\end{figure}

The stellar mass is a derived quantity, dependent on a number of assumptions relating the directly-observed stellar light to an implied stellar mass. This in turn requires a model of the observed surface brightness from which a mass can be derived. In this work, we use the surface-brightness model of NGC 3115 presented in \cite{emsellem99}, which used the multi-Gaussian Expansion (MGE) technique \citep{monnet92, emsellem94}. MGE models fit a series of \(2{\rm D}\) Gaussians to the observed photometric isophotes. One advantage of using the MGE approach is that for any inclination that is trialled by the dynamical model, the Gaussians have an analytic deprojection into an intrinsic \(3{\rm D}\) model. This results in a fast (though not necessarily unique) description of the mass for that inclination, forming the framework within which the dynamical model is computed. \cite{emsellem99} used a combination of high-resolution HST and ground-based photometry to compute a photometric fit out to \(\sim 300\si{\arcsecond}\), which is necessary in order to ensure that the stellar mass model comfortably encloses the extent of such a nearby galaxy. The collapsed data cube and MGE surface brightness model are shown in \cref*{img:mge}.\par
\begin{figure}
    \centerline{
        \includegraphics[width=\columnwidth]{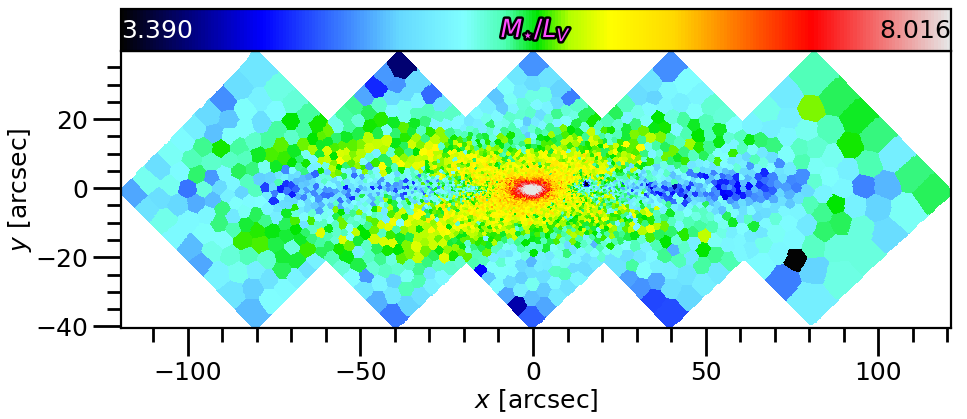}
    }
    \caption{A map of \(M_\star/L\) in \(V\)-band inferred from full spectral fitting with variable age and metallicity, and assuming a Kroupa IMF.}
    \label{img:sml}
\end{figure}

Typical \shw\ model implementations, and many dynamical models in general, assume that the conversion from light to mass can be done with a single global scale for a given galaxy; that is, a spatially-constant stellar mass-to-light ratio (\(M_\star/L\)). This implies that the observed light originates from only a single population of stars. In our work, however, we consider the interplay between dynamical and stellar-population structures, and so the spatial variations in the \(M_\star/L\) are of particular importance. Since we have already characterised the presence of multiple stellar populations within this galaxy in \cref*{sec:srsfh}, it is therefore possible to incorporate the spatial structures in the stellar populations to obtain a more accurate conversion from light to mass. It is imperative that we attempt to construct an accurate {\em input} mass model that takes into account this information in order to maintain self-consistency when we analyse the outputs. There have already been recent implementations of this for other purely-dynamical modelling techniques by parametrising the derived \(M_\star/L\) map into a \(1{\rm D}\) radial profile, which is then used to scale the surface-brightness model accordingly \citep{poci17, mitzkus17}. We compute a \(V\)-band \(M_\star/L_V\) map for NGC 3115 based on the {\sc ppxf} fits from \cref*{sec:srsfh}, shown in \cref*{img:sml}. \(V\)-band is consistent with both the spectral range used in the MUSE observations and the original photometry used by \cite{emsellem99}.
It is immediately clear, however, that the \(M_\star/L_V\) map of NGC 3115 can not be well-approximated by a radial profile due to the `lobe' features along the major axis created by a relatively young stellar disk. In order to derive an accurate stellar-mass model, we take a different approach here in order to maintain the \(2{\rm D}\) information from the SFH. We scale the surface-brightness MGE (\mgeS) by the MUSE \(M_\star/L_V\) map directly in order to obtain a mass `image', to which we fit a new {\em mass} density MGE (\mgeT).\par
\begin{figure}
    \centerline{
        \includegraphics[width=\columnwidth]{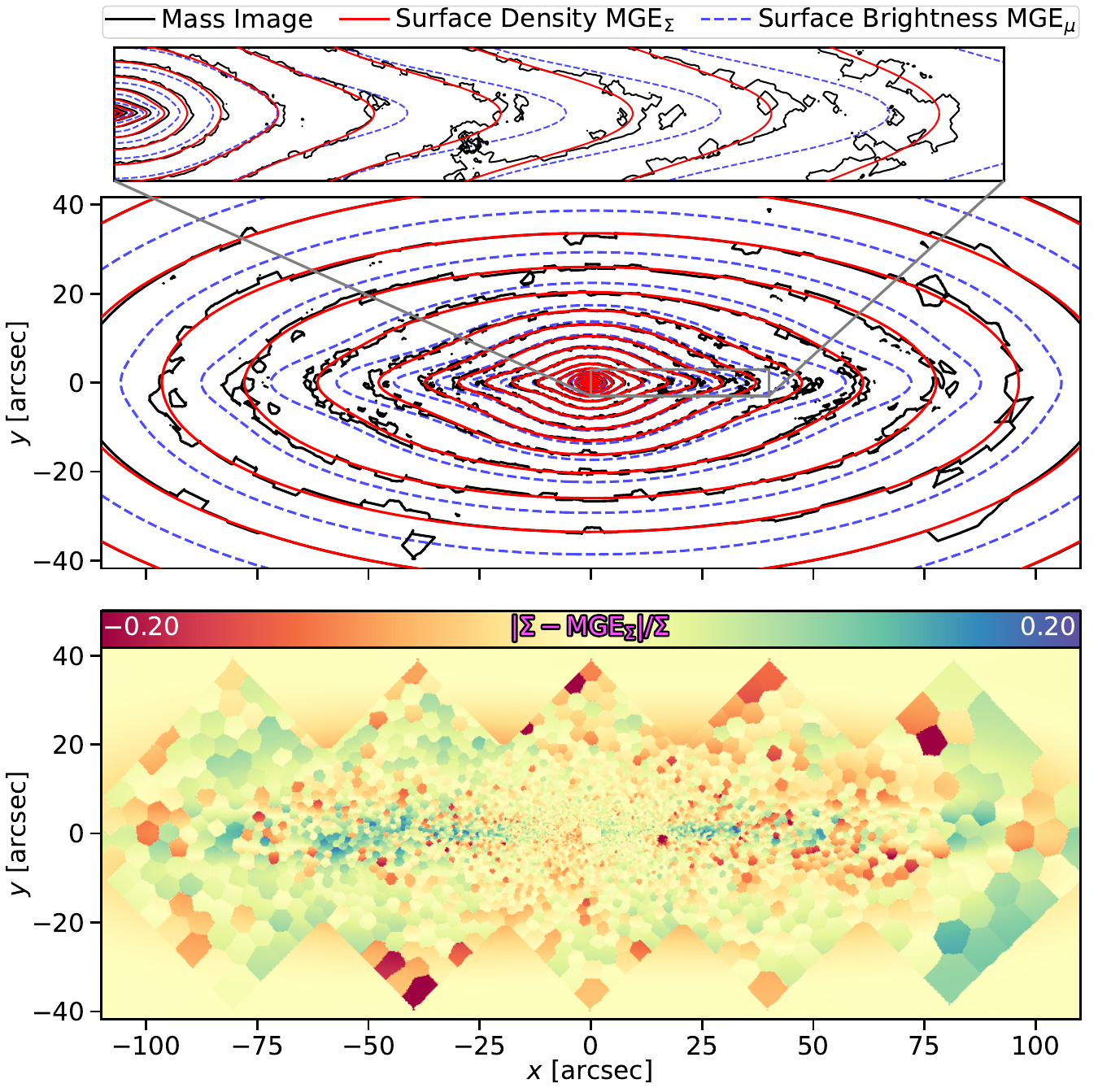}
    }
    \caption{The fit in red to the iso-mass contours of the scaled mass image in black (top). The surface-brightness MGE is shown for reference in dashed blue. The panel shows the FOV of the IFU, while the zoom-in shows only the disk region between \(0\) and \(40\si{\arcsecond}\). The fractional residuals between the mass `image' and \mgeT\ are shown in the bottom panel.}
    \label{img:MGES}
\end{figure}

One issue when comparing photometric and IFU observations is the difference in the size of the FOV. To overcome this, we first evaluate \mgeS\ on an image grid that is sufficiently large \citep[comparable to the FOV used in][]{emsellem99}. We then cast the \(M_\star/L_V\) map onto the same image scale. To populate the pixels that lie outside of the MUSE FOV, we assume that the \(M_\star/L_V\) is constant at large radius. This already appears to be the case in the MUSE observations for \(R\gtrsim 80\si{\arcsecond}\), and so the exterior pixels are fixed to the average value of the outermost bins of \cref*{img:sml}. \cref*{img:MGES} presents \mgeS\ and \mgeT\ on the approximate scale of the stellar disk, which clearly highlights the impact of considering the \(2{\rm D}\) \(M_\star/L_V\) map.
The \mgeT\ has rounder iso-mass contours (compared to \mgeS) in the region where relatively young stars contribute a lot of luminosity but not much mass. Assuming a constant \(M_\star/L\) would have attributed too much mass to the disk region (along the major axis), which in turn would have biased the dynamical model and the inferred stellar populations. \mgeT\ is tabulated in \cref*{app:MGET}\footref{fn:exData}.

\subsubsection{Dark-Mass Model}
To include contributions from non-luminous mass, we include a parametrisation of the dark matter (DM) halo by assuming it has the form of a generalised spherical Navarro, Frenk, and White \citep[NFW;][]{navarro96} halo, as described in Eq. (3) of \cite{zhu18b}, but included here for completeness in \cref*{eq:gnfw}.
\begin{flalign}\label{eq:gnfw}
    M_{\rm DM} (r) &= M_{200} \cdot g\left(C_{\rm DM}\right) \left[ \ln\left(1 + \frac{rC_{\rm DM}}{r_{200}}\right) - \frac{\frac{rC_{\rm DM}}{r_{200}}}{1+ \frac{rC_{\rm DM}}{r_{200}}} \right]&&&&
\end{flalign}
where \(g\left(C_{\rm DM}\right) = \left[\ln\left(1+C_{\rm DM}\right) - \frac{C_{\rm DM}}{1+C_{\rm DM}}\right]^{-1}\), and \(M_{200} = 200\times\frac{4}{3} \pi \rho_c r_{200}^3\). Here, \(\rho_c= \frac{3 H^2}{8 \pi G}\) is the critical density of the Universe with Hubble constant \(H\) and gravitational constant \(G\). While we do not test different dark matter prescriptions in this work, we note that the broader method described in this work is not tied to the specific parametrisation of the dark matter halo.\par
The central SMBH similarly contributes non-luminous mass to the mass model, and therefore affects which orbits exist within the gravitational potential. Our models include a spatially-localised non-luminous Plummer potential \citep{dejonghe87} to represent the mass of the SMBH, \(M_{\bullet}\), defined as
\begin{equation}
\Phi(r) = - \frac{G M_\bullet}{\sqrt{r^2 + a^2}}
\end{equation}
where \(a\) is the Plummer core radius (effectively the SMBH softening length). In the region \(r \leq a\), the gravitational potential of the model is dominated by the Plummer potential. We fix \(a=\SI{0.008}{\arcsecond}\), which is below the MUSE resolution, in order to avoid numerical issues when integrating orbits close to the SMBH.\par
\subsection{Orbit Solution}\label{ssec:orbSol}
The \shw\ code generates a library of orbits that are physically permitted for the given mass model. In our model, we sample the orbits over 42 logarithmically-spaced radial starting locations, which conserve the first integral of motion; the energy. There are also 15 linearly-spaced locations in \(I_2\), the second conserved integral of motion. Finally, we sample the third (non-classical) conserved integral of motion, \(I_3\), from 12 linearly-spaced starting locations. We may cheaply (without re-integrating) increase the size of the orbit library, by simply mirroring orbits about their integrals of motion to allow for counter-rotation. In the general triaxial code, this occurs only for those orbits which are sampled on the \((x,z)\) meridional plane, doubling the number of those orbits. Moreover, to boost the number of box orbits accessible to the model, we launch a new set of box-only orbits, adding another factor to the orbit library \citep{vandenbosch08}. Therefore, the full number of orbits in the model is \(N_{\rm orb.} = 3 \times N_E \times N_{I_2} \times N_{I_3} = 3 \times 42 \times 15 \times 12 = 22680\). In order to minimise the discreteness of the orbits, each starting position is `dithered' \citep[see][]{cappellari06} in the three-dimensional integral space by a factor of 5, thereby sub-sampling each  \((E, I_2, I_3)\) location into a grid of \(5^3\) adjacent starting positions. This high sampling ensures that the small physical scale of the observation is probed sufficiently, while also covering a large enough volume to describe the full extent of the galaxy. Compared to the extensive work done on CALIFA data \citep{zhu18b}, our data has \(5\) times higher spatial resolution, and our orbit library is slightly over \(5\) times larger. Specifically, we can estimate the approximate density of orbits across our FOV. By taking the conservative approach of considering the orbital starting locations, we find an average of \(\sim 102\ \si{orbits/arcsec^2}\). From this density of orbits, we conclude the orbit library contains sufficient resolution to accommodate our data. Each of these orbits is integrated for 200 complete periods.\par
With a complete orbit library, the \shw\ code then solves for the weighted linear combination of these orbits that best reproduces the (projected) observations. The orbital weights are computed by a Non-Negative Least-Squares (NNLS) fit to all observed constraints simultaneously, producing a weight per orbit, per spatial bin. For a given combination of orbits, the combined model is projected and convolved with the PSF to mimic the observations, binned onto the same Voronoi bins, and finally, the intrinsic LOSVD of the model are compared to the measured kinematic moments in each spatial bin. It is clear here that each kinematic moment extracted from the observed spectra therefore provides additional constraining power, in order to discriminate between models with similar lower-order kinematic moments.\par
While the orbit {\em integration} is carried out in the gravitational potential (mass model), the orbit {\em solution} is fit to the observed luminosity-weighted properties - namely, the surface brightness MGE, and all 6 kinematic moments. The linear combination of orbits is therefore luminosity-weighted also, consistent with the observations and the SFH described in \cref*{sec:srsfh}.
\subsection{Model Free Parameters}\label{ssec:mfp}
The process described in \cref*{ssec:orbSol} is for the single fixed gravitational potential within which the orbits were integrated. While the \emph{projected} mass model is fixed by \mgeT, the intrinsic galaxy shape that gives rise to this projection remains unknown. It is therefore necessary to optimise the free parameters of the intrinsic gravitational potential in order to obtain not only a best-fitting set of orbits for a given potential, but also the best-fitting potential itself. The parameters of the gravitational potential are described here.\par
For a triaxial system, the baryonic component can be described by three intrinsic shape parameters, \((q, p, u)\). These intrinsic shapes are translated into projected viewing angles according to Eqs. (7)\(-\)(9) of \cite{vandenbosch08}. While we fit for \(q\) and \(p\), we fix the third intrinsic shape parameter to \(u=0.99999\). This imposes the assumption that the shape of the baryonic component of the galaxy does not depend on the azimuthal angle at which we are observing it (that is, invariant under rotations about the rotation axis). This is a reasonable assumption in our case, where we are modelling a nearly-edge-on, approximately axisymmetric galaxy.\par
There are also free parameters of the dark mass contributions to the gravitational potential. From \cref*{eq:gnfw}, our model gains two free parameters; the concentration of dark matter \(C_{\rm DM}\), and the fraction of dark-to-stellar mass within \(r_{200}\), \(f_{\rm DM}\left(r_{200}\right) = M_{200} \big/ M_\star\). We also vary \(M_\bullet\) in order to accurately probe the orbits in the central region of the galaxy.\par
The final free parameter of the \shw\ model is the \(M_{\rm dyn.}/L\) scaling, which we refer to as \(\Upsilon\) to avoid confusion with the SFH-derived \(M_\star/L_V\). Although we have incorporated the shape of the stellar component into the gravitational potential using the MUSE \(M_\star/L_V\) map, this final free parameter is a global (spatially-constant) scale of the total mass. It allows the model to produce a deeper or more shallow potential, as required by the dynamics, due to possible systematics in the derived \(M_\star/L_V\) including the assumption about the IMF, as well as the accuracy of the dark matter parametrisation. Since \(\Upsilon\) is a global scaling, the shape of the gravitational potential does not change. Therefore, this parameter can be efficiently optimised by scaling all of the orbital velocities by \(\sqrt{\Upsilon}\), and refitting the model. A value of \(\Upsilon=1\) implies that the input gravitational potential, in terms of the the absolute stellar \(M_\star/L\), assumed \(M_{\bullet}\), and assumed DM parametrisation, contains the correct enclosed mass to reproduce the observed kinematics.\par
We are thus left with 6 free parameters for the \shw\ model, described in \cref*{tab:FP}.
\begin{table*}
\centerline{
\begin{tabu}{c|c|c|c|c}
    {\bf Parameter} & {\bf Description} & \(\pmb{\Delta}\) & {\bf Start} & {\bf Best}\\\hline\hline
    \(m_{\rm BH}\) & Black-Hole Mass & \(0.75^*\) & \(1.000 \times 10^9\) & \(1.185\times 10^9\ \si{\Msun}\)\\\hline
    \(q\) & Intrinsic Shape & \(0.005\) & \(0.001\) & \(3.500\times 10^{-4}\)\\
    \(p\) & Intrinsic Shape & \(0.005\) & \(0.999\) & \(0.979\) \\
    \(u\) & Intrinsic Shape & \(-\) & \(-\) & \(0.999\) \\\hline
    \(\theta\) & Viewing Angle & \(-\) & \(-\) & \(85.200\si{\degree}\)\\
    \(\phi\) & Viewing Angle & \(-\) & \(-\) & \(88.743\si{\degree}\) \\
    \(\psi\) & Viewing Angle & \(-\) & \(-\) & \(90.004\si{\degree}\) \\\hline
    \(C_{\rm DM}\) & DM Concentration & \(0.250\) & \(15.500\) & \(15.250\) \\\hline
    \(\log_{10}\left[f_{\rm DM}\left(r_{200}\right)\right]\) & DM Fraction at \(r_{200}\) & \(0.050\) & \(2.000\) & \(1.950\) \\\hline
    \(\Upsilon\) & Global \(M/L\) & \(0.005\) & \(1.000\) & \(0.980\ \si{\Msun/\Lsun}\)
\end{tabu}
}
\caption{The free parameters of the \shw\ model, their corresponding step sizes, \(\Delta\), and the final best-fitting values. \(u\) is fixed in our model, while \(\theta, \phi, \psi\) are derived from \(q,p,u\). These parameters therefore have no \(\Delta\). \(^*\) Note that the black-hole mass is sampled logarithmically, and so its \(\Delta\) is multiplicative rather than additive.}
\label{tab:FP}
\end{table*}
To find the global best-fitting model, we employ the following \(\chi^2\) grid search. From a starting guess for each parameter, we consider those locations which are \(\pm 5\Delta\) (the step size defined in \cref*{tab:FP}) away. This is to ensure that a sufficient volume of the parameter-space is probed to avoid local minima. Including the central position, there are then 3 initial trials for each free parameter. In total, for \(N\) {\em gravitational} free parameters (excluding \(\Upsilon\)), the model begins with \(3^N\) trials. Each of these \(3^5 = 243\) initial gravitational-potential models is evaluated for 3 values of \(\Upsilon\). Once these have completed, the best-fitting location becomes the new `centre', and each free parameter is explored in increments of \(\pm \Delta\) in a similar manner. This is repeated until a location is found such that all surrounding models in the \(N{\rm D}\) space produce worse fits to the data, which is followed by one final run with \(\pm 0.5\Delta\) in order to accurately characterise the best-fitting region.\par
\begin{figure}
    \centerline{
        \includegraphics[width=\columnwidth]{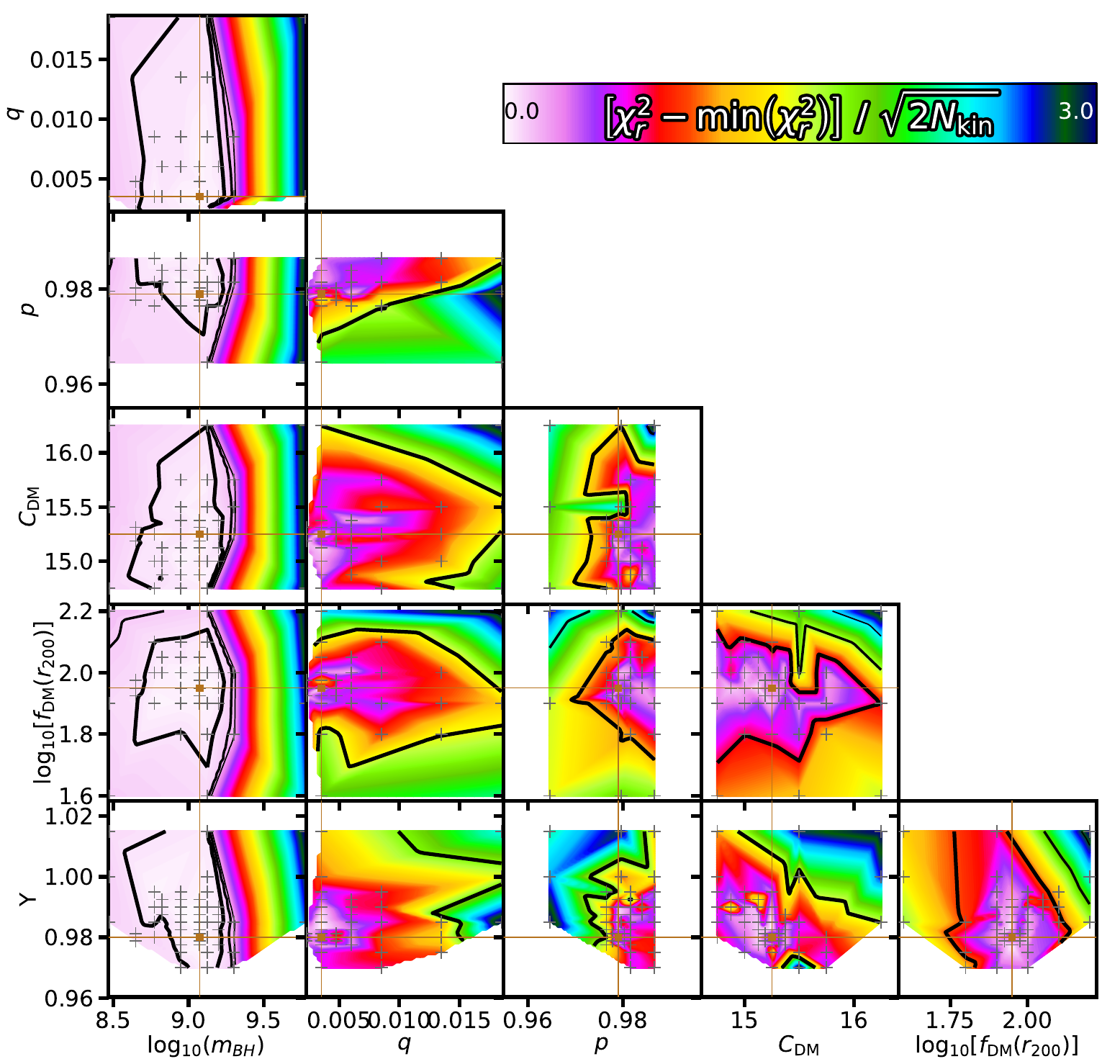}
    }
    \caption{The grid search over the free parameters of the \shw\ model. Each panel shows the marginalised \(\Delta\chi^2\) contour surface in colour. The black solid lines show, from bold to thin, the \(1-\), \(2\)-, and \(3-\sigma\) ranges, respectively. The grey `\(+\)' symbols show the underlying grid of models that were generated, and the magenta lines indicate the final best-fitting values.}
    \label{img:corn}
\end{figure}
\begin{figure*}
    \centerline{
        \includegraphics[width=1.05\textwidth]{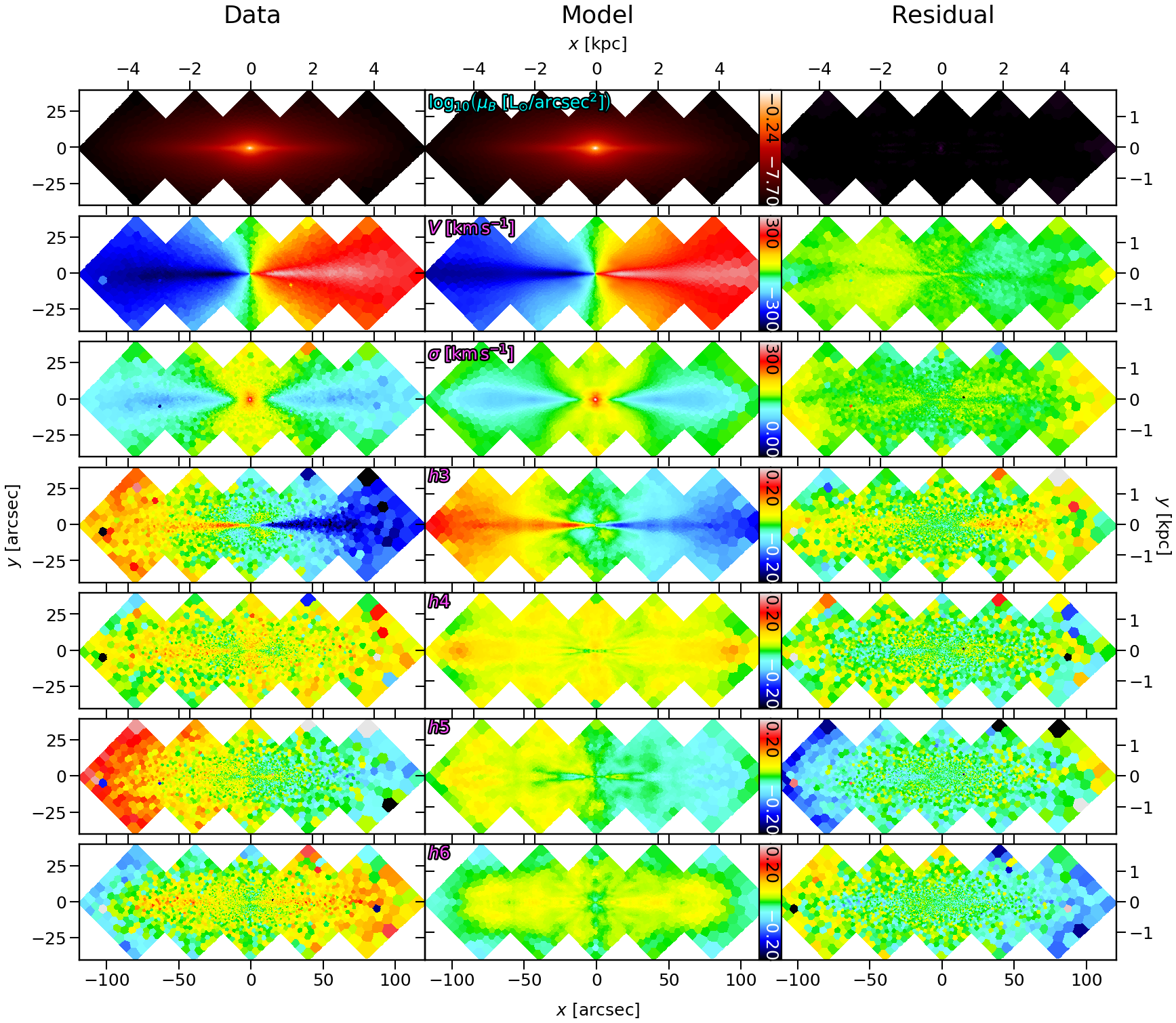}
    }
    \caption{The data (left), model (center), and residuals (right) for the best-fitting \shw\ model of NGC 3115. From top to bottom (labelled in the top-left corner of each model panel) is the surface brightness, followed by all kinematic moments. The model is fit to the surface brightness MGE, rather than the surface brightness directly. The residuals for the surface brightness have been offset such that \(0\) is the black end of its corresponding colourbar. All other residual panels have been offset such that \(0\) corresponds to the green (centre) of their corresponding colourbars.}
    \label{img:shw}
\end{figure*}

The metric for model comparison used in this work follows the formalism of \cite{zhu18b}. That work introduced a normalised \(\chi^2\) metric in order to account for spatial pixels which are not truly independent. It is defined as
\begin{flalign}
\chi_r^2 &= \chi_{\rm kin}^2 \cdot N_{\rm kin} \big/ {\rm min}\left(\chi_{\rm kin}^2\right) &&&&
\intertext{with}
\chi_{\rm kin}^2 &= \sum\limits_{g=1}^{N_{\rm GH}}\sum\limits_{a=1}^{N_{\rm aper}} \left(\frac{d_{a,g} - m_{a,g}}{e_{a,g}}\right)^2 &&&&\tag*{}\\
\intertext{for the data, model, and error of the \(a\)-th aperture of the \(g\)-th Gauss-Hermite kinematic coefficient, \(d_{a,g}\), \(m_{a,g}\), and \(e_{a,g}\), respectively, and}
N_{\rm kin} &= N_{\rm GH} \times N_{\rm aper} &&&&\tag*{}\\
&= 6 \times 4881 &&&&\tag*{}\\
&= 29286 &&&&\tag*{}
\end{flalign}
for the number of Gauss-Hermite coefficients fit in the \shw\ model, \(N_{\rm GH}\), and the number of spatial Voronoi apertures, \(N_{\rm aper}\). In this way, a standard deviation of this \(\chi_r^2\) distribution is given by \(\sqrt{2 N_{\rm kin}}\), and the grid search for NGC 3115 is illustrated in \cref*{img:corn}.
The corresponding best-fitting \shw\ model is shown in \cref*{img:shw} for all seven projected constraints.

\subsection{Intrinsic Angular Momentum}\label{sec:intrAM}
Following the optimisation of the gravitational potential parameters, the best-fitting \shw\ model provides intrinsic information about the galaxy, instead of being typically limited to projected quantities. The orbits of the model have intrinsic \(3{\rm{D}}\) angular momentum and orbital anisotropy, and the model has true LOSVD rather than parametrisations of these distributions. These intrinsic properties are a key requirement for the application of our methodology.\par
One such property that we utilise directly here is the intrinsic angular momentum. Each orbit has a corresponding angular momentum vector \(\vec{L} = (L_x, L_y, L_z)\). We consider a scaled version of this intrinsic angular momentum, the `circularity', which was introduced by \cite{zhu18b} and is defined as
\begin{flalign}
    \lambda_z &= \overline{L_z} \Large/ \left( \overline{r} \cdot \overline{V_2} \right) &&&&\\
    \intertext{with}\\
    \overline{L_z} &= \overline{x V_y - y V_x} &&&&\tag*{}\\
    \overline{r} &= \sqrt{\overline{x^2 + y^2 + z^2}} &&&&\tag*{}\\
    \overline{V_2}^2 &= \overline{V_x^2 + V_y^2 + V_z^2 + 2 V_x V_y + 2 V_x V_z + 2 V_y V_z} &&&&\tag*{}
\end{flalign}
We use the bar notation to denote that these values are averaged over all of the \((x, y, z, V_x, V_y, V_z)\) points along a given orbit's integrated path. The probability distribution of the \(\lambda_z-r\) plane for the best-fitting \shw\ model is shown in \cref*{img:lzCut} (see \cref*{sssec:dynSel}). It shows only the region constrained by the kinematics, even though the model orbits can extend beyond this in order to fit the full stellar MGE at large radius.\par
Circularity is a projection of the full \(3{\rm D}\) orbital distribution, and `clumps' in this projection identify families of orbits with similar orbital characteristics in the meridional plane. This definition separates circular and box/radial orbits, having circularity values of \(\left|\lambda_z\right|=1\) and \(\lambda_z=0\), respectively. It is analogous to the `stellar-spin' parameter \citep[defined in][]{emsellem07} in that it is a metric to discriminate between rotation- and dispersion-supported systems (or orbits).

\subsection{Applying Stellar Populations to the Orbital Structures}\label{ssec:dynComps}
At this point, we have a dynamical model, fitted to the observed kinematics, that provides information on the orbital distribution function by way of orbital luminosity weights. We now wish to couple this information to the observed stellar population properties, in terms of the (luminosity-weighted) mean age and metallicity maps. We do this by assigning individual ages and metallicities to the fitted orbital components, noting that a `component' can be composed of a collection of orbits from our dynamical model. We first consider how such components may be defined from the model, and then describe how we associate them with stellar population information to fit the observations.
\subsubsection{Dynamical Selection}\label{sssec:dynSel}
The \shw\ model provides a full description of the orbital structure of our galaxy, being a combination of many hundreds of distinct orbital families. However, in terms of coherent galaxy sub-structures, we may expect far fewer components. In our own Milky Way, for instance, we broadly recognise a `thin disk', `thick disk', `bulge', and `halo'. Indeed, as described in the introduction, such broad component definitions are often applied when considering the decomposition of external galaxies.  By combining both stellar kinematics and populations, we here seek to test such `component' concepts.\par
As described above, circularity phase-space gives a simple projection of the full orbital phase-space, from which broadly-distinct orbits can be tracked as a function of radius in the galaxy. We take advantage of this as a way of potentially reducing the number of distinct orbital families by instead grouping them to form `components'. This brings computational advantages by reducing the dimensionality of the problem, as well as giving a clear definition of what the (dynamically, chemically, and chronologically) distinct components are. We trial a number of conceptually-different approaches here, and compare their results.
\begin{description}
\item[{\bf Sample I:} {\it Conventional Decomposition}]
We refer to Sample I as a `conventional' decomposition because we define bins in circularity that closely approximate classic galactic components; namely, a thin disk, thick disk, and bulge. We use the ranges defined by \cite{zhu18b}; \(\lambda_z >0.8\), \(0.25 < \lambda_z \leq 0.8\), and \(\lambda_z \leq 0.25\), representing the thin disk, thick disk, and bulge, respectively.
\item[{\bf Sample II:} {\it Small Sampling in Circularity}]
In this instance, we group the orbits into small bins of \(\lambda_z\). We do not assign physical meaning to the resulting `components', but merely investigate how the extra freedom affects the stellar-population fitting.
\item[{\bf Sample III:} {\it Small Sampling in Circularity and Radius}]
This approach samples the \(2{\rm D}\) circularity space in small bins of circularity and radius. Once again, we avoid attributing physical meaning to these components, but investigate specifically if radial changes (compared to \(\lambda_z\) sampling only) can improve the fit significantly. For consistency, the \(\lambda_z\) bins are the same as Sample II. Also, to align with the sampling of the underlying orbits from the \shw\ model, the sampling of the circularity phase-space in radius is logarithmic in the inner region (but linear outside the FOV to avoid overly-large bins).
\item[{\bf Sample IV:} {\it Orbit-Based Decomposition}]
Finally, in the limit of increasingly smaller bins in circularity, each individual orbit may be considered as a separate component. Dealing directly with orbits is conceptually a more robust implementation, as it avoids issues with the `projection' of the intrinsic \(3{\rm D}\) orbital phase-space into the circularity domain. It is, however, computationally expensive, and can result in an ill-conditioned problem for large numbers of orbits.
\end{description}
For each criterion, we consider only those components/orbits that are given non-zero weight in the best-fitting \shw\ model, as all others do not contribute any mass to the model.\par
\begin{figure}
    \centerline{
        \includegraphics[width=\columnwidth]{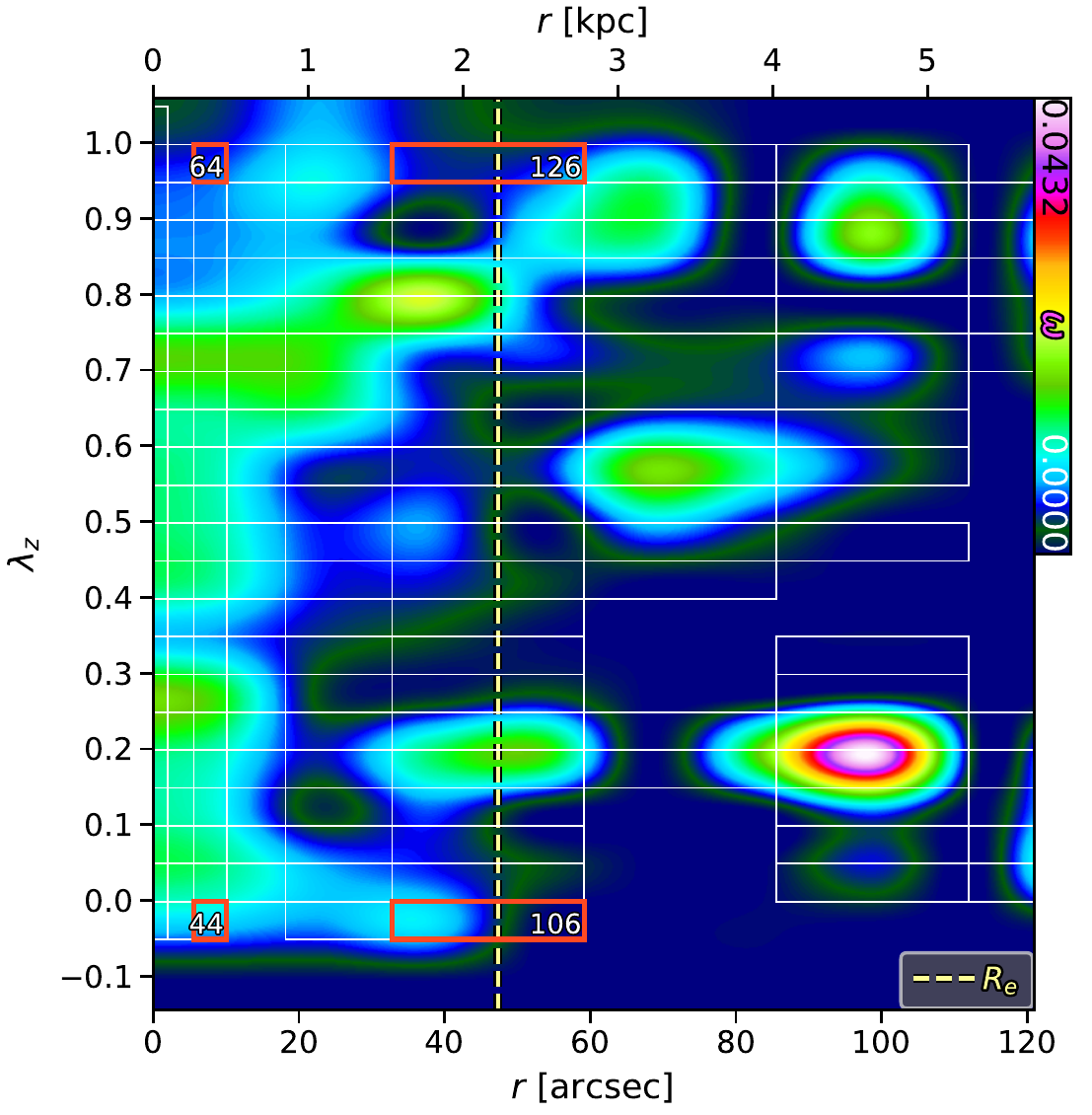}
    }
    \caption{The \(\lambda_z-r\) phase-space of the best-fitting \shw\ model of NGC 3115. The density map is the orbital probability distribution (circularity; \protect\cref*{sec:intrAM}), with the dynamical component selections of Sample III shown in white boxes. The vertical dashed line marks \(R_e\). Highlighted and numbered are components representative of different regions of the circularity phase-space; see \protect\cref*{img:decomp127}.}
    \label{img:lzCut}
\end{figure}
\begin{figure*}
    \centerline{
        \includegraphics[width=1.05\textwidth]{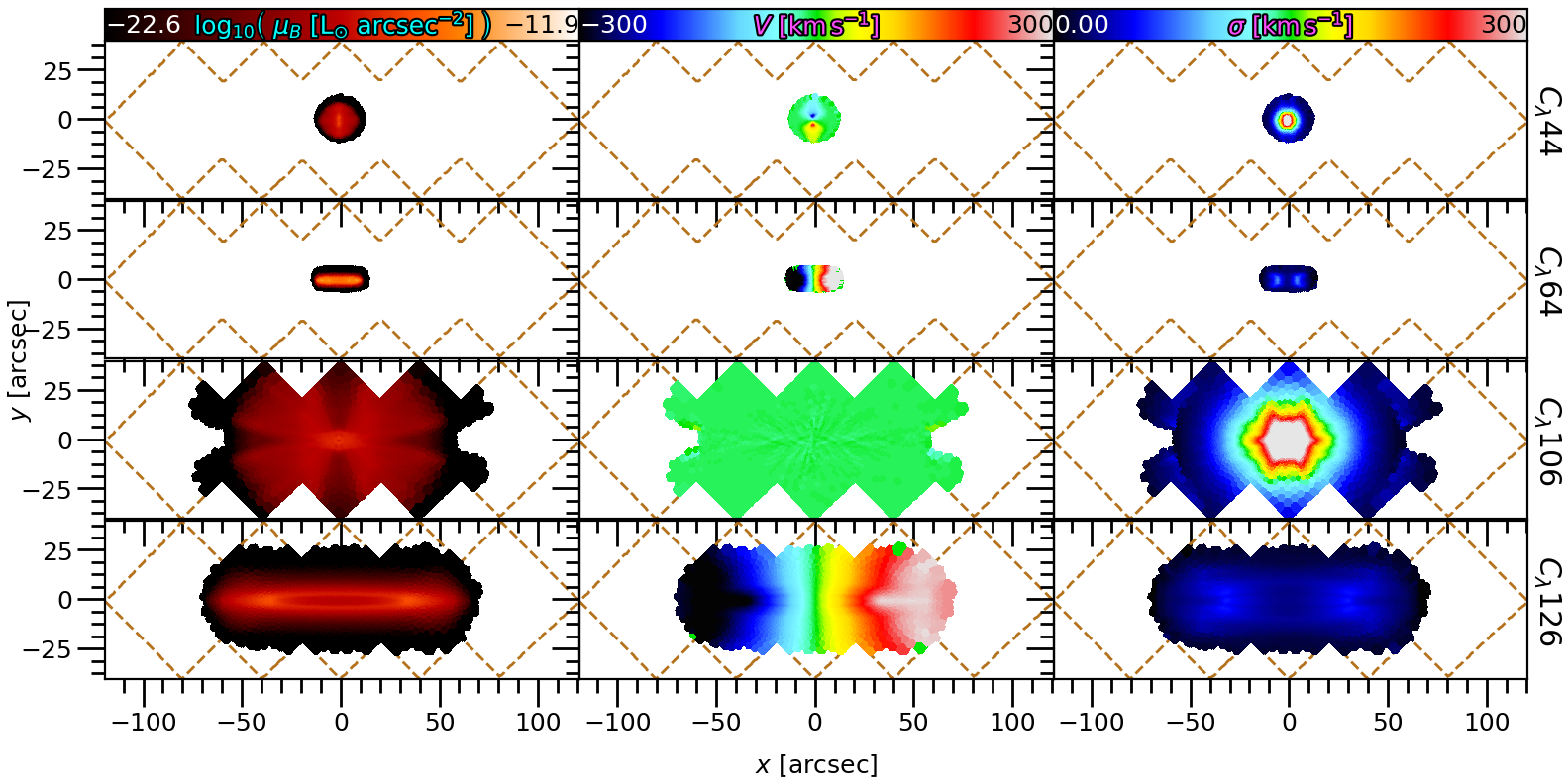}
    }
    \caption{The surface brightness (left), mean velocity (middle), and velocity dispersion (right) of four characteristic dynamical components. Each component is presented on the same binning and FOV as the data for consistency. The FOV is outlined by the brown dashed line, and the component labels are on the far right of each row (refer to \protect\cref*{img:lzCut}). Note that each of these components does \emph{not} contribute equal mass to the dynamical model, but were selected to illustrate different regions in circularity phase-space.}
    \label{img:decomp127}
\end{figure*}
\begin{figure}
    \centerline{
        \includegraphics[width=1\columnwidth]{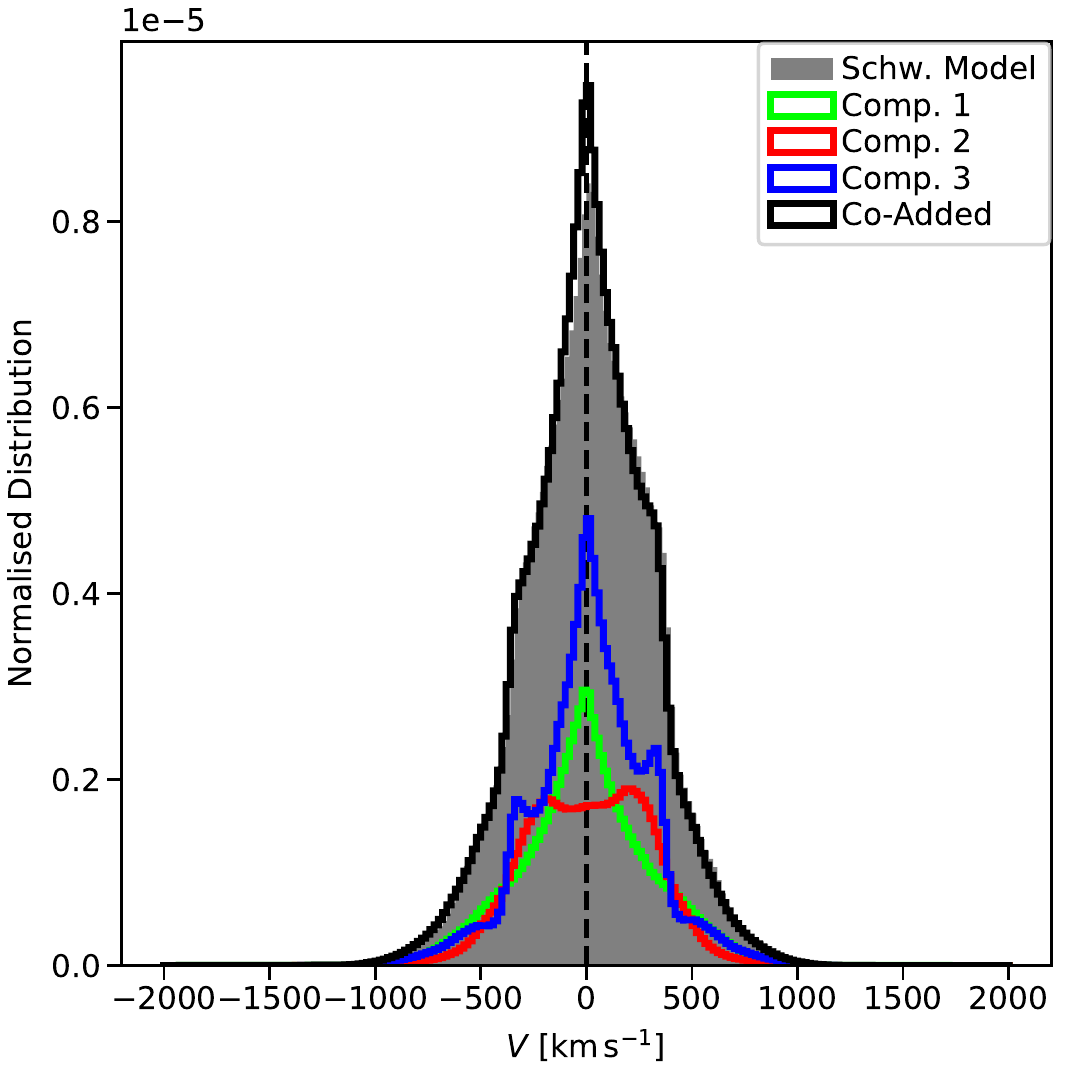}
    }
    \caption{The LOSVD in a single aperture, for the dynamical components of Sample I (coloured histograms), their sum (black), and from the original \shw\ model (grey filled).}
    \label{img:losvd}
\end{figure}

We refer to Sample III as the illustrative case for the remainder of this work (see \cref*{ssec:numbComp}), and \cref*{img:lzCut} shows the corresponding components in circularity phase-space. The gaps in the grid correspond to `components' that have zero weight.
By the nature of the \shw\ model, we can investigate these dynamical components in greater detail, in order to gauge their spatial extent and physical properties. \cref*{img:decomp127} shows the surface brightness and kinematics of a subset of characteristic dynamical components, selected to contrast the different regions in the circularity phase-space.
These properties are derived as statistical moments of the intrinsic LOSVD of each component - the integral, mean, and dispersion for the surface brightness, mean velocity and velocity dispersion, respectively. The dynamical decomposition effectively divides the LOSVD in each aperture amongst the resulting components. Each component's LOSVD, therefore, need not be Gaussian. This is illustrated in \cref*{img:losvd}, using the Sample I decomposition for clarity, and is why we restrict our visualisation of the components to the low-order kinematics.
Nevertheless, it is clear from \cref*{img:decomp127} that the lowest \(\lambda_z\) components are truly pressure-supported, while the highest \(\lambda_z\) components are rotation-supported.

\subsubsection{Dynamics/Stellar Populations Associations}\label{sssec:fsksp}
The \shw\ model provides the luminosity-weights per aperture (Voronoi bin), \emph{for each component}. The SFH provides the mean luminosity-weighted stellar-population properties per aperture. We can therefore assign such properties to the dynamical components, by equating their luminosity-weighted average with the SFH values. For a single aperture, this can be expressed as
\begin{flalign}\label{eq:row}
    \Phi^{\rm SFH} &= \frac{\sum_{j=1}^{N_{\rm comp.}} \omega^j \phi^j}{\sum_{j=1}^{N_{\rm comp.}} \omega^j} &&&&
\end{flalign}
where \(\Phi^{\rm SFH}\) is the luminosity-weighted stellar-population value (\(t\) or \(Z\)) from the SFH, \(\omega^j\) is the orbital weight of the \(j\)-th dynamical component, and \(\phi^j\) is the unknown stellar-population value (\(t\) or \(Z\)) of the \(j\)-th dynamical component. If we normalise the orbital weights beforehand, such that
\begin{flalign}
    \tilde{\omega}^j &= \frac{\omega^j}{\sum_{j=1}^{N_{\rm comp.}} \omega^j}\ , &&&&
\end{flalign}
we can then express the entire problem, over all apertures, as a matrix equation
\begin{flalign}\label{eq:matrix}
    \begin{pmatrix}
        \tilde{\omega}_1^1             & \tilde{\omega}_1^2             & \cdots & \tilde{\omega}_1^{N_{\rm comp.}} \\
        \tilde{\omega}_2^1             & \tilde{\omega}_2^2             & \cdots & \tilde{\omega}_2^{N_{\rm comp.}} \\
        \vdots                 & \vdots                 & \ddots & \vdots \\
        \tilde{\omega}_{N_{\rm aper}}^1 & \tilde{\omega}_{N_{\rm aper}}^2 & \cdots & \tilde{\omega}_{N_{\rm aper}}^{N_{\rm comp.}}
    \end{pmatrix} \cdot \begin{pmatrix}
        \phi^1 \\ \phi^2 \\ \vdots \\ \phi^{N_{\rm comp.}}
    \end{pmatrix} &= \begin{pmatrix}
        \Phi^{\rm SFH}_1 \\
        \Phi^{\rm SFH}_2 \\
        \vdots \\
        \Phi^{\rm SFH}_{N_{\rm aper}}
    \end{pmatrix} &&&&
\end{flalign}
where one row in the matrix (superscripts) denotes the weights for all dynamical components in a single aperture - equivalent to \cref*{eq:row} - and one column (subscripts) corresponds to the weights of a single dynamical component in all apertures. We now simply require a linear (computationally-efficient) matrix inversion in order to find the \(\vec{\phi}\), which thereby associates the dynamically-identified components with individual stellar populations. For \(\phi\) and \(\Phi\) in this work, we investigate both age and metallicity. Unlike other similar methods recently proposed in the literature \citep{long18}, our method does explicitly allow consideration of the \(3{\rm D}\) distribution of stellar populations in projection, as the age and metallicity parametrisation explicitly adds linearly (unlike the line strength indices considered in that work). The matrix inversion is done in the framework of a Bounded-Value Least-Squares (BVLS) fit, which is akin to NNLS problems, but rather than a positivity constraint, the solution bounds are specified explicitly. Here we impose the boundary values of the SSP library in order to maintain consistency with our SFH. This is not necessary in general - for instance, if \(\vec{\Phi}\) is generated without the use of an SSP library - and we emphasise that the best-fitting \(\vec{\phi}\) is continuous within the bounds and not tied to the specific sampling of any SSP library. While there exists mild covariances between the stellar age and metallicity, our spectral fits from \cref*{sec:srsfh} employ a small linear regularisation. This regularisation assumes a smooth SFH in both age and metallicity in order to break the degeneracy. Moreover, in the subsequent fitting described in this section, our model needs to reproduce the spatial structure in the SFH maps, not just the age and metallicity values of an individual bin. This combination of spatial and temporal coherence allows our model to mitigate effects due to covariances. We therefore treat age and metallicity as orthogonal parameters, and simply fit them independently.\par
Note that the different decomposition criteria described in \cref*{sssec:dynSel} are merely re-distributions of the weights \(\tilde{\omega}^{j}\) into different numbers of \(N_{\rm comp.}\), while \cref*{eq:matrix} and the method itself remain unchanged.

\section{Results}\label{sec:results}
Here we present the main results of this work. We first present the various combinations of dynamical components and stellar population properties, and then how these combinations lead to inferences of the formation history of this galaxy.
\subsection{Constraining the Required Number of Dynamical Components}\label{ssec:numbComp}
\begin{figure*}
    \centerline{
        \includegraphics[width=1.05\textwidth]{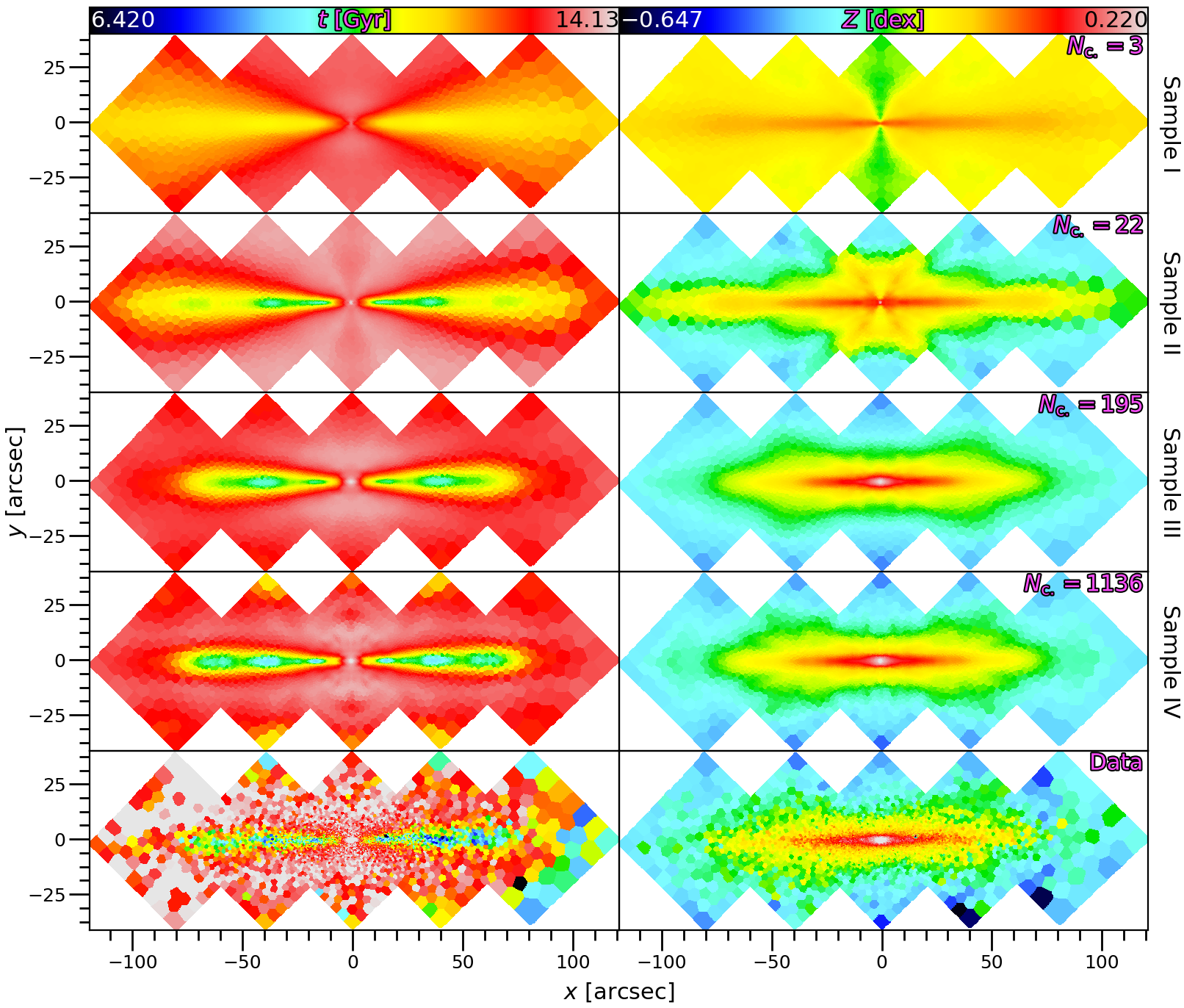}
    }
    \caption{The luminosity-weighted age (left) and metallicity (right) maps. The first four rows are for the different dynamical selection criteria (\protect\cref*{sssec:dynSel}), labelled on the far right. The number of dynamical components resulting from each sample is inset in the top right of each metallicity panel. The bottom row shows the maps derived from the SFH (\protect\cref*{sec:srsfh}).}
    \label{img:mwCompOrb}
\end{figure*}

\cref*{img:mwCompOrb} shows the fits to the measured luminosity-weighted mean stellar age and metallicity for all of the sample criteria tested in this work (see \cref*{sssec:dynSel}).
It is immediately clear that the conventional few-component decompositions are completely unable to reproduce the structure in the stellar populations. This implies a rather dramatic disconnect between the photometric, kinematic, and chemical structures seen in the galaxy, at least from the point of view of a handful of components. However, we note that while the surface brightness of the components from Sample I follow typical S\'ersic-law expectations, the assumption that they have a single monolithic formation is strongly inconsistent with the data. This suggests that studies assuming the conventional S\'ersic approach and fitting a few components should include stellar population {\em gradients}, rather than a single population per component as done here, even for evolved objects like NGC 3115.\par
Similarly, it appears that regardless of the degree of freedom in \(\lambda_z\) as in Sample II, the fits are unable to reproduce the SFH, in particular the metallicity. This reaffirms the existence of radial gradients in stellar populations, with these gradients being stronger in metallicity, as has been seen by many studies previously both in general \citep[for example, see][]{sarzi18, parikh18, poetrodjojo18} and specifically for NGC 3115 \citep[for example, see][]{strom76,pastorello14}. However, we can additionally conclude here that these gradients must exist within each component \emph{even at fixed \(\lambda_z\)}.\par
Unsurprisingly, as the number of degrees of freedom (dynamical components) increases, so too does the quality of fit. Surprisingly, though, once the circularity phase-space is well-sampled in both dimensions as in Sample III, there already appears to be sufficient freedom in the model to reproduce the stellar-population maps, without having to consider every individual orbit. This is not necessarily expected {\em a priori}, but allows us to dramatically reduce the size of the parameter-space for subsequent analyses by using Sample III as the default dynamical decomposition moving forward. This is because, while there is a factor of 10 increase in \(N_{\rm comp.}\) between Sample III and IV, their fits to the stellar-population maps are almost indistinguishable.\par
The fit described in \cref*{sssec:fsksp} is done on each bin of the observations. However, due to the fact that orbits from the \shw\ model overlap in projection, multiple orbits will contribute to the mean age and metallicity of a single bin. There are necessarily different weighted mixtures of ages and metallicities that give the same average values, producing a degeneracy when associating stellar-population properties to dynamical components. One option would be to reduce the number of components until the fit is adversely affected, however this requires arbitrary choices of which components to remove. A more natural approach is to apply linear regularisation. The regularisation scheme used here prefers solutions where adjacent dynamical components/orbits have similar stellar population properties given an otherwise degenerate alternative. It is described in detail in \cref*{app:regul}.

\subsection{Recovering the Dynamical SFH}
\begin{figure*}
    \centerline{
        \includegraphics[width=1.05\textwidth]{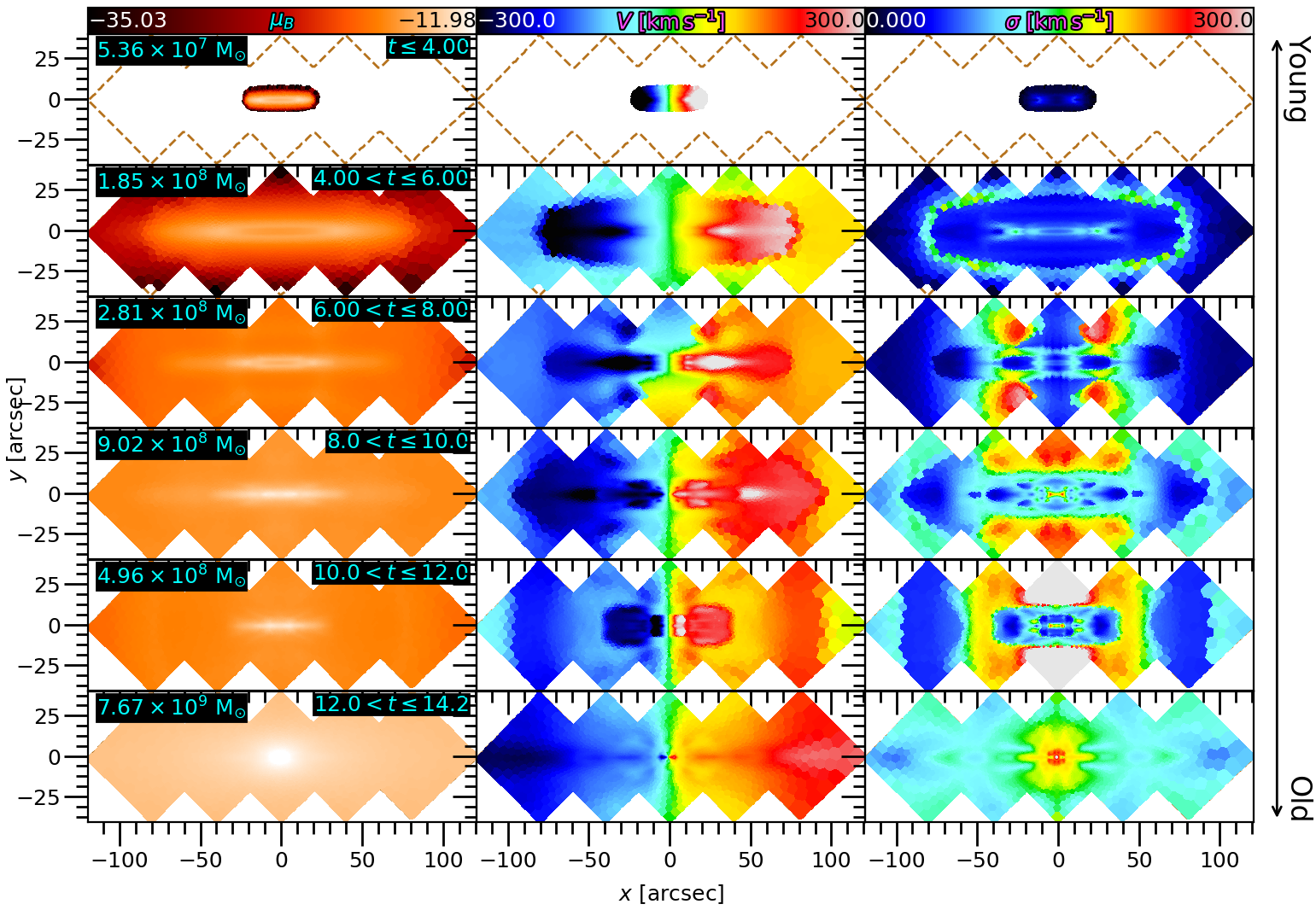}
    }
    \caption{The surface brightness (left), mean velocity (middle) and velocity dispersion (right) of the dynamically-selected components, binned into their corresponding mean stellar age from young (top) to old (bottom). The bin ranges and corresponding stellar masses of each bin are inset in the upper right and left of each surface brightness panel, respectively.}
    \label{img:sfhAge}
\end{figure*}

\begin{figure*}
    \centerline{
        \includegraphics[width=1.05\textwidth]{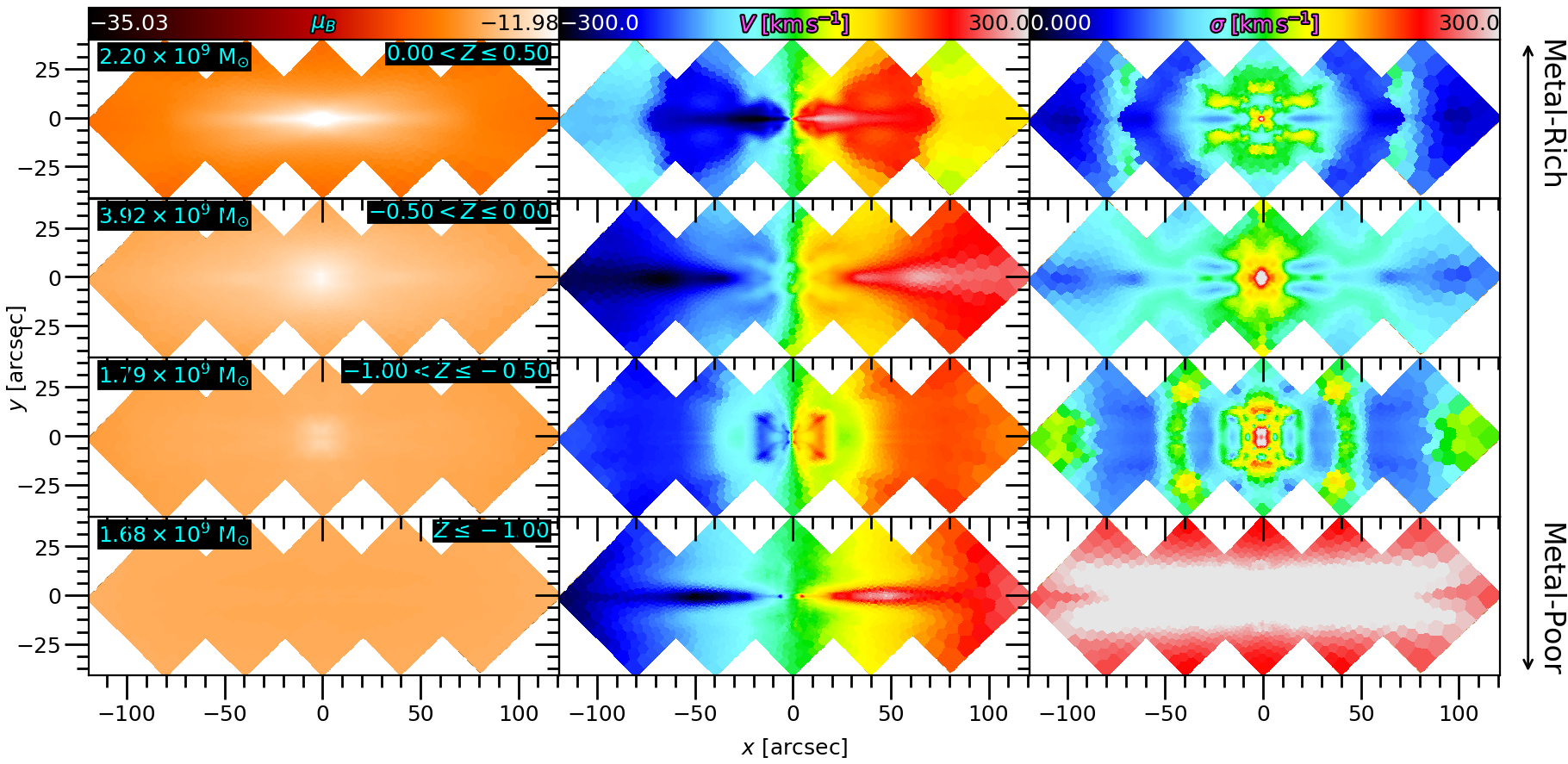}
    }
    \caption{As \protect\cref*{img:sfhAge}, but for the stellar metallicity.}
    \label{img:sfhMetal}
\end{figure*}

Following these fits and adopting the results from Sample III, we now have a set of components for NGC 3115 with both fitted kinematics and stellar populations. We have already been able to investigate the structures in components of the galaxy based on a dynamical selection criterion (see \cref*{img:decomp127}). Now, however, we can investigate NGC 3115 from an orthogonal perspective: the nature of spatial structures in components selected on their stellar populations. This produces spatially-resolved maps for components of the galaxy in bins of age and metallicity, which is effectively a conventional SFH, except that it originates from the spatial distribution of {\em dynamical} components. A conventional full-spectral-fitting SFH was computed in \cite{guerou16}. For comparison, we construct the same bins in age and metallicity as that work, which are shown in \cref*{img:sfhAge,img:sfhMetal}, respectively.
By considering for now just the left columns of \cref*{img:sfhAge,img:sfhMetal}, it is clear that the bulk of the stellar mass is old, and in a spheroidal structure concentrated at the centre but extending across the FOV. We see a small portion of relatively young stars, that exist in a much more flattened structure along the plane of the galaxy. There is remarkable agreement between the purely spectral analysis in \cite{guerou16} and our combined approach, despite the fundamental differences between the two methodologies. Moreover, from the metallicity panels, we find a central elongated metal-rich component. This is surrounded by a near-solar diffuse component, which is slightly overdense in the centre. We find a portion of the mass in an even more diffuse metal-poor halo-like component. The absolute masses presented in each panel of \cref*{img:sfhAge,img:sfhMetal} are computed as the fractional mass (determined by the \shw\ model) of \mgeT, integrated within the MUSE FOV. This ensures that these stellar mass measurements include the information from the stellar \(M_\star/L_V\) map, and are consistent with the dynamical model and the spectral SFH.\par
In addition to the conventional SFH analysis possible with this method, we can extract {\em intrinsic} kinematics for each of these age and metallicity bins. This is possible at present uniquely from our combined methodology. It allows us to investigate the chemistry and kinematics simultaneously and self-consistently, in a spatially-resolved manner. These kinematics are shown in the middle and right columns of \cref*{img:sfhAge,img:sfhMetal}. As mentioned in \cref*{sssec:dynSel}, the decomposition of the full \shw\ model by either dynamical (that section) or population (this section) properties necessarily divides up the LOSVD in each aperture. This is why the kinematics in \cref*{img:sfhAge,img:sfhMetal} may appear non-physical and discontinuous. These components do not exist in isolation, and the kinematics of each component are physically meaningful only in the context of the full dynamical model. Moreover, the bins in age and metallicity are discrete, further contributing to the discontinuities in the kinematic maps. However, visualising the results in this way is useful because it allows us to investigate the {\em relative} dynamical properties of the different stellar populations in a completely spatially-resolved manner. It is in fact quite clear from these figures that, for instance, the youngest ages are heavily rotationally-supported, and that the most metal-poor components are highly pressure-supported - findings that we expand on in the following sections.

\subsection{Intrinsic Stellar Velocity Dispersion and Stellar Populations}
We may delve further into the model by investigating its intrinsic properties, and any direct correlations between the dynamical and stellar-population properties. To this end, we study how the properties of the intrinsic velocity ellipsoid and the assigned stellar populations are related, focusing on the vertical velocity dispersion in order to isolate features in the disk plane. We present in this section the results of this investigation, based on the intrinsic properties derived from the \shw\ model and the subsequent population fits.\par
\begin{figure*}
    \centerline{
        \includegraphics[width=\textwidth]{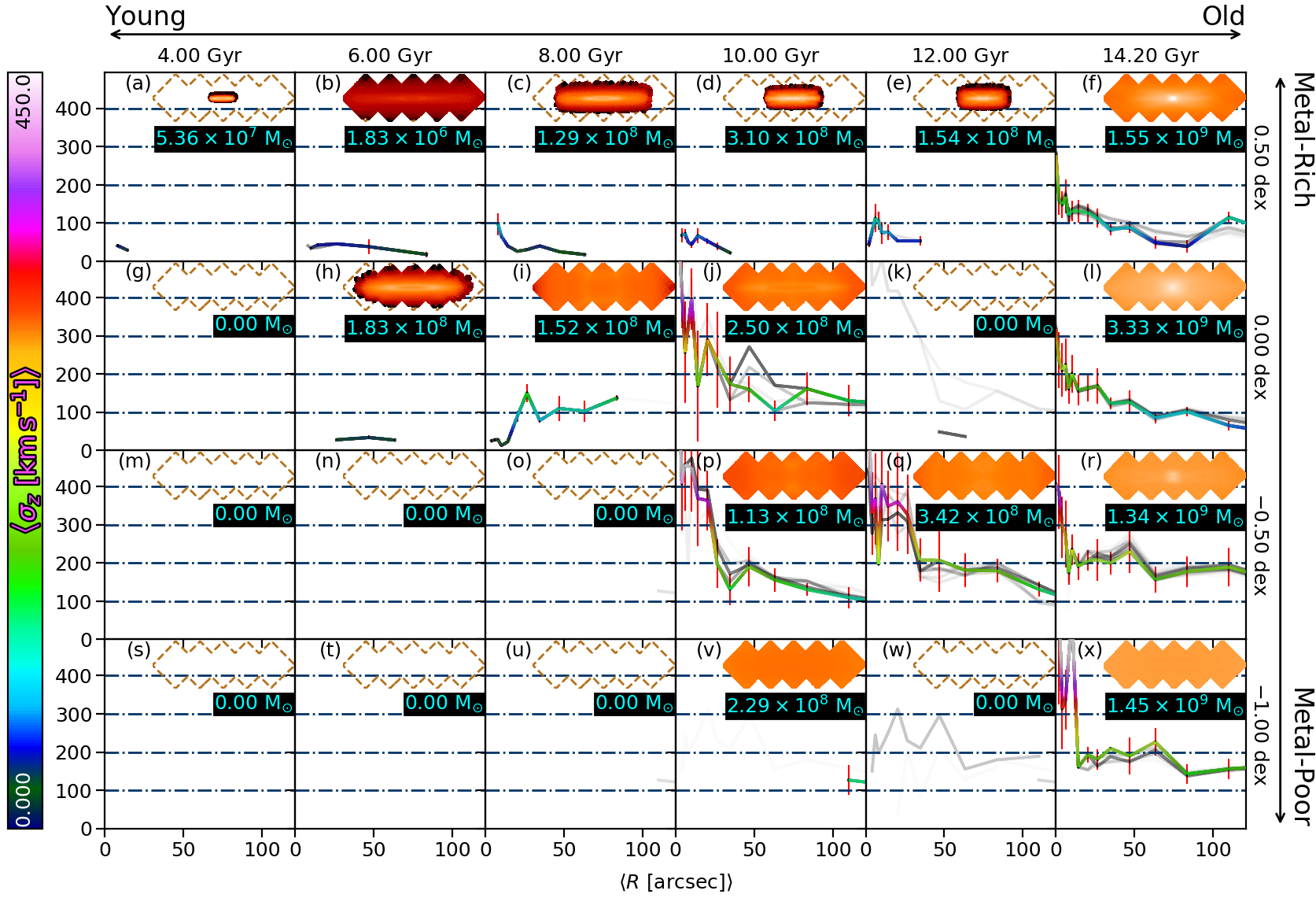}
    }
    \caption{{\em Main}: The radial profiles of the \(z\)-component of the {\em intrinsic} velocity dispersion for the dynamical components grouped by their mean stellar-population properties. Each curve is coloured by its \(\sigma_z\) values to allow an easier quantitative comparison between panels. The variance within the annuli (see text) are shown as errorbars. The grey curves are Monte Carlo simulations of the stellar-population fits. They are computed by randomly perturbing the stellar-population maps from the spectral SFH (\protect\cref*{sec:srsfh}) within their measurement uncertainties, and re-fitting to the dynamical components. Each trial is plotted with high transparency, such that only the regions with high density of curves are visible. We note that, in a few instances, the Monte Carlo simulation has populated distinct panels from the best-fit model. {\em Inset}: The surface-brightness of each age/metallicity bin, observed at the best-fitting projection of the \shw\ model \((\sim 85\si{\degree})\) on the same FOV as the MUSE observations. All surface-brightness panels are on the same colour-scale. The absolute stellar masses are inset in cyan.}
    \label{img:2dDispR}
\end{figure*}

We consider the \(z\)-component of the velocity dispersion, \(\sigma_z\), in order to probe the structure and properties of the disk region of NGC 3115. As we have the complete phase space information for every orbit in our model, we can compute the intrinsic moments of the orbital motions. This includes the first `true' moment of the velocity distribution along the three principle axes \((V_x, V_y, V_z)\), the second moments, and all the cross-terms. The second moment of the velocity distribution can be approximated as
\begin{flalign}
\langle V_k^2 \rangle &= \sigma_k^2 + \langle V_k \rangle^2 &&&&
\end{flalign}
for first and second moments \(\langle V_k \rangle\) and \(\langle V_k^2 \rangle\), respectively, and velocity dispersion \(\sigma_k\), where \(k \in [x, y, z]\) is the Cartesian axis. Therefore, to compute the velocity dispersion, it follows trivially that
\begin{flalign}\label{eq:vd}
\sigma_k &= \sqrt{\langle V_k^2 \rangle - \langle V_k \rangle^2} &&&&
\end{flalign}
The outputs of the \shw\ model are first transformed from Cartesian to cylindrical coordinates. To construct radial profiles, we transform the logarithmic energy sampling from the original \shw\ model from spherical coordinates to cylindrical coordinates. This produces a cylindrical radius equivalent to the original sampling in \(E\). For clarity, we combine all the orbits within one PSF of the galaxy centre and treat this as a single position, as we cannot directly resolve this region. Radial profiles of the first and second moments - \(\left\langle V_z \right\rangle (R)\) and \(\left\langle V_z^2 \right\rangle (R)\), respectively - are constructed by computing the luminosity-weighted sum of the orbits' moments that lie within radial annuli. These are then converted into \(\sigma_z(R)\) with \cref*{eq:vd}. Finally, all components belonging to a particular age and metallicity bin are averaged to produce stellar-population-selected velocity-dispersion profiles. These are shown in \cref*{img:2dDispR}.
The radial extent of each profile is determined by two factors; whether there is mass at a given radius from a given age/metallicity bin (that is, the spatial extent of the corresponding surface-brightness distribution), and whether \cref*{eq:vd} is numerically defined at a given location in a given bin, which depends on its particular combined kinematics from the dynamical model. We present the marginalised plots (over age and metallicity) in \cref*{app:sigma_z}

\subsection{Modelling Systematics}\label{ssec:modSys}
Before interpreting the wealth of information in \cref*{img:2dDispR}, we briefly discuss here the possible sources of systematic uncertainties arising from this method, and how reliable the subsequent formation history is.\par
In all of the dynamical samples presented in \cref*{sssec:dynSel}, the components are defined by hard rectangular grids in the orbital probability distribution. It is possible that this is responsible for some of the substructure that can be seen in some panels of \cref*{img:decomp127}. Some components appear to be a superposition of underlying dynamical structures, indicating that those components could somehow be further divided until unique dynamical features are isolated. It may be possible to mitigate these effects to some extent by introducing a more complex method for grouping dynamically-similar orbits to define the components, but such a method would have to remain contiguous in the circularity phase-space in order to conserve the mass of the dynamical model. Testing and implementing a diverse range of methods for the dynamical selection is beyond the scope of this work.\par
By construction, each row of the matrix in \cref*{eq:matrix} - each spatial bin of the observations - is assigned the same set of \(\vec{\phi}\). The consequence of this is that each component is mono-age and mono-metallicity. For the triaxial \shw\ model, each component is also at least point-symmetric. Therefore, the resulting fit in \cref*{img:mwCompOrb} can not reproduce any asymmetry in the SFH maps, whether that asymmetry is physical or otherwise. There is mild (non-physical) asymmetry in the SFH maps for this particular data-set. The effect of this would be a slightly larger discrepancy between the data and model maps in \cref*{img:mwCompOrb} due to the implicit averaging between the \(\pm x\) sides. While we mitigate the most significant deviations by masking the right-most bins during the fit, there is still a gradual sky residual gradient across the FOV. This can be further seen in the youngest ages of the disk on the left and right sides, which would contribute to the uncertainties from that fit. However, such an issue is specific to the data-set in use here, and general applications of our method would not be subject to these uncertainties.\par
In \cref*{ssec:numbComp,app:regul}, we address the possibility that the solutions to \cref*{eq:matrix} may be unstable/degenerate. To estimate what impact this degeneracy may have specifically on our interpretation of \cref*{img:2dDispR}, we run 100 Monte Carlo simulations by perturbing the measured SFH maps within their uncertainties, and re-fitting the dynamical components as described in \cref*{sssec:fsksp}. Running through the entire analysis, we generate a new set of \(\sigma_z(R)\) curves for each trial, and these are plotted within each panel of \cref*{img:2dDispR}. It is clear from the dense clustering of curves that the solutions are reasonably stable. It is possible for some components to traverse the bins in stellar populations, but this is only because their best-fit age and metallicity are close to the boundaries of the bins, while the absolute change in the individual ages and metallicities remains small.\par
More broadly, in order to accurately estimate the possible systematic uncertainties associated with a method of this nature, it is necessary to run our analysis on mock data derived from high-resolution hydrodynamical galaxy simulations. This effort is considerable in and of itself, and has been in preparation in parallel with our application here to real data (Zhu et al., in prep.). That work will test the accuracy and reliability of the recovery of stellar population properties for dynamically-selected components. However, in this work specifically, there is such clear structure in the stellar populations and kinematics due to the proximity of the galaxy and the quality of the data, that this method can be applied with confidence in its ability to recover the underlying properties of the galaxy.

\section{Discussion}\label{sec:formation}
\subsection{Galactic Components of NGC 3115}
The many facets of our comprehensive model can be seen concisely in \cref*{img:2dDispR}, in which we study the galaxy across stellar age and metallicity, and simultaneously with a metric for the spatially-resolved, intrinsic dynamical properties. We characterise these galactic features here.
\begin{description}[itemsep=1em]
\item[{\it Old, Metal-Rich, Compact, Hot Spheroidal Bulge}] Firstly, we find a bulge-like component at the oldest age, with a mild spread in metallicity. This is clear from the central, compact spheroidal peak in surface brightness in the three most metal-rich panels at \(14.2\ \si{Gyr}\); panels \((f)\), \((l)\), and \((r)\). This is accompanied by a corresponding central \((R \lesssim 15\si{\arcsecond})\) peak in \(\sigma_z\) in these bins \((\sigma_z \gtrsim 100\ \si{km\ s^{-1}})\), indicating that this region is pressure-supported. This component has only undergone mild chemical enrichment, likely due to secular stellar-evolution processes, with no other significant evolution - neither in its kinematics nor populations. By making the appropriate cuts in age, metallicity, circularity, and radius in order to isolate only the bulge contributions to the relevant panels, we estimate that this component contains \(\sim 3.9 \times 10^{9}\ \si{\Msun}\).\par
\item[{\it Metal-Rich, Extended, Cold Disk}] We also find evidence for a disk-like structure that is enriched, and present at {\em all} ages. This is clear from the surface brightness in panels \((a)-(f)\), and also panel \((h)\), which all show elongated structures in the plane of the galaxy. Again, this is corroborated by the dynamics, which show that this component has \(\sigma_z \lesssim 50\ \si{km\ s^{-1}}\) at intermediate radii where the disk dominates, implying that it is supported primarily by coherent rotation (cold orbits). In particular, even panel \((f)\) has a signature of the disk in both the surface brightness and the velocity dispersion. This is an interesting finding, as a galactic component that is simultaneously cold, old, and enriched would be sensitive to any dynamical perturbations over the galaxy's history. Therefore, the mere existence of such an old population in a disk configuration immediately implies a fairly quiescent history for NGC 3115. We see that star formation in the disk continued smoothly from the earliest times, gradually declining until it ceased \(\sim 4\ \si{Gyr}\) ago. The resulting stars are naturally enriched, and are progressively more dynamically cold - a progression that we quantify further in \cref*{ssec:resolve}. In our model, the disk component is \(\sim 1.6 \times 10^{10}\ \si{\Msun}\).
\item[{\it Old, Metal-Poor, Diffuse, Hot Stellar Halo/Thick Disk}] Finally, we find a diffuse halo/thick-disk-like component that is dynamically hot (\(\sigma_z \gtrsim 100\ \si{km\ s^{-1}}\), including at large radius), metal-poor \((\nuclide{[Z/H]} \lesssim -0.50\ \si{dex})\), and mostly old \((t \gtrsim 10\ \si{Gyr})\). This can be seen in panels \((p)-(x)\), which all have a similar featureless spatial distribution, and relatively high \(\sigma_z\). Panels \((j)\) and \((l)\) appear to have contributions from this component as well, but superimposed with contributions from the disk and bulge, respectively. This stellar halo/thick-disk component contains \(\sim 4.2 \times 10^{10}\ \si{\Msun}\).
\end{description}

\subsection{Implications for the Formation History of NGC 3115}\label{ssec:formation}
The current paradigm for galaxy formation, in particular of ETG, details the hierarchical assembly of massive galaxies via the merging of smaller systems, which result in expectedly pressure-supported systems due to the nature of the merging process \protect\citep[for instance, see][and references therein for a general review]{somerville15}. Merging during the lifetime of a galaxy is a key aspect, and often the focus, of many cosmological and/or hydrodynamical simulations \citep[for example, see][]{bird13,athanassoula16,elichemoral18}. Moreover, merging is often invoked to explain many phenomena seen in observational studies \citep[for example, see][]{arnold11,lidman13,guerou16}. However, these mergers are typically assumed to have been of two (or more) fully-formed progenitor galaxies, implying that sufficient cosmological time had passed for such galaxies to first form, then merge. As discussed above, we find a very old and dynamically cold population in the disk of NGC 3115, implying that any destructive major merger had to have occurred prior to the formation of this disk. \cite{laurikainen13} arrived at the same conclusion when they detected cold `lenses' \citep[an elliptical structure with a sharp inner edge in surface brightness;][]{kormendy79} within \SZ\ galaxies that were very old - also \(\sim 14\ \si{Gyr}\). The Universe had, at such early times, considerably different physical conditions to the environments in which the simulated progenitor galaxies were formed. For instance, the progenitor galaxies in \cite{athanassoula16} were explicitly modelled after ``nearby galaxies''. As a result, such scenarios may not appropriate for our inferences here, and we therefore look for a possible alternate formation mechanism to explain the results we see in \cref*{img:2dDispR}.\par
One such mechanism is `compaction' \citep{dekel09,zolotov15} at early times. Compaction involves the rapid dissipative contraction of highly-perturbed gas into the central regions of galaxies in the form of cold streams. These streams can trigger star-formation in this region, which is reportedly quenched rapidly for massive galaxies \citep{zolotov15}. Therefore, compaction could also explain the enrichment gradient in the central bulge component, which is all at early times. Moreover, since the gas cooling is dissipative, and the cold streams are not as disruptive as major mergers, this method of formation could also explain the persistence of the old cold disk structure \citep{dekel09}, even if the streams entered after the old disk formed. The remaining gas which had been accreted via the cold streams would go on to form the main stellar disk at progressively younger ages until the gas reservoir is exhausted.\par
Irrespective of the formation mechanism for the central bulge and main disk, the outer stellar halo/thick-disk component is strongly consistent with an accreted origin, mainly from dry minor mergers. This is due to the combination of old ages, low metallicities, pressure-supported kinematics, and featureless yet extended surface brightness. This component forms a significant portion \((\sim 0.68 M_\star)\) of the stellar mass of NGC 3115. Interestingly, the cosmological simulations of \cite{oser10} predict that for `intermediate mass' galaxies such as NGC 3115, the accreted material constitutes \(\sim 65\%\) of the stellar mass at \(z=0\), which is remarkably consistent with our model. Given such a significant contribution to the stellar mass budget, persistent minor mergers could also explain how NGC 3115 transformed into an \SZ\ galaxy, by building up the `thick' disk and diluting any spiral arms that may have been present in the progenitor object. While previous works have claimed that either environmental perturbations \citep{bekki11}, internal disk instabilities \citep{saha18}, or major mergers \citep{querejeta15,tapia17,diaz18,frasermckelvie18} are the likely formation paths for \SZ\ morphologies, our model is inconsistent with these mechanisms. NGC 3115 is a field \SZ, making environmental perturbations unlikely, and both internal disk instabilities and major mergers would likely destroy the old disk structure that we find in our model.

\subsection{Comparison to Previous Inferences}
Due to its proximity, NGC 3115 has been studied widely in the literature, using many techniques that produce orthogonal constraints on the inferred formation history. We investigate here how our results compare with these works.\par
\cite{arnold11} use Supreme-Cam imaging and long-slit GC observations to study the kinematic and metallicity properties of NGC 3115. They found that an early violent major merger would explain the relatively high rotation and flattening of the central bulge component. \cite{guerou16} proposed that a small number of progenitors with \(\log_{10}(M_\star/\Msun) \sim 10\) is consistent with both the surviving angular momentum of this central component, as well as its enrichment at such early times. \par
\cite{guerou16} suggested that some portion of the gas reservoir that was present during the major merger survived the interaction, went on to cool and eventually form the dynamically cold, younger disk stars. Our results indicate that the cooling of the gas and the formation of new stars happens `immediately' (within the oldest age bin), and stars continued to form on progressively colder orbits through to the \(4\ \si{Gyr}\) bin, until this gas was consumed. Their interpretation of the excess gas does indeed agree with our results, however our model does not need to invoke a major merger as the source, and an alternate origin for the gas could be the cold streams associated with an early compaction phase.\par
\cite{arnold11} claim that steeper metallicity gradients can also be the result of passive accretion of low-mass, low-metallicity satellites which lower the average metallicity in the outskirts, thereby steepening the gradient. This accretion is also consistent with the observed metallicity map for NGC 3115. \cite{brodie12} and \cite{cantiello14} find strong evidence that the observed colour bimodality in the GC population in NGC 3115 is driven by an underlying metallicity bimodality. This supports not just an accreted origin for the blue population of GC, but specifically accretion from dry minor-mergers, locking in the low metallicity of the in-falling objects. Our metal-poor, dynamically-hot, diffuse components also strongly favour passive accretion. This is because low-mass objects would be relatively metal-poor, and many such in-falling objects would impact at arbitrary angles imparting no net angular momentum, while increasing the \(\sigma_z\).

\subsection{Resolved Studies and Galaxy Formation Simulations}\label{ssec:resolve}
By exploiting the intrinsic, spatially-resolved properties of our model, we can begin to compare directly to results from resolved studies of the Milky-Way and Local Group, as well as cosmological hydrodynamical galaxy formation simulations. This allows us to leverage the look-back capabilities of the simulations as well as the resolving power of the local observations in order to strengthen the interpretation of our results. Moreover, by extending the intrinsic correlations seen in the Milky Way to external galaxies, we begin to gauge how unique the Milky Way is.\par
Age-velocity dispersion relations have been found in many resolved \citep{nordstrom04,grieves18} and simulated \citep{martig14b} studies. They primarily conclude that stars born earlier have a higher velocity dispersion compared to those formed later. The underlying physical cause of this relation, however, has evaded a general consensus in the literature. Some scenarios broadly claim that the increase in velocity dispersion is a dynamical effect of internal interactions that build up over time, with a number of different specific mechanisms being proposed as the culprit. Since the older stars have been experiencing these interactions for a longer period of time, they should therefore show the largest increase in velocity dispersion. Evidence in favour of such internal mechanisms has come from both observations \citep{yu18} and simulations \citep{saha10,grand16,aumer16}. An alternative explanation is that in the early Universe, conditions were generally more chaotic \citep{wisnioski15}, so that any stars born at that time were more likely to have higher velocity dispersion. As conditions gradually settled over cosmic time, stars were being born in progressively lower-dispersion conditions. A number of studies have identified the conditions at birth as the dominant effect in determining a population's present-day velocity dispersion, again both from resolved observations \citep{leaman17} and simulations \citep{bird13,ma17}. Finally, the comparison between observations and simulations in \cite{pinna18} has identified the underlying complexity and inherent degeneracy in discriminating between these scenarios. They claim that many of the effects described above likely play a role to a varying degree, and that the imprint of some mechanisms fade over time, further complicating any attempt to constrain the physical cause of disk heating.\par
\begin{figure}
    \centerline{
        \includegraphics[width=1.05\columnwidth]{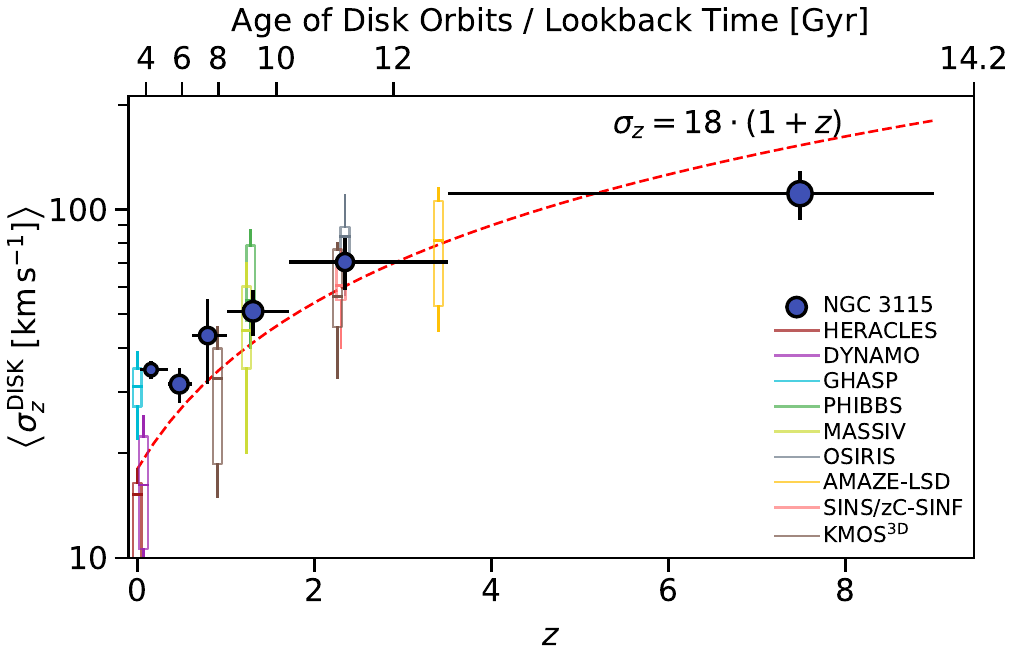}
    }
    \caption{The vertical component of the intrinsic velocity dispersion of the orbits associated with the disk region (see text) of NGC 3115 (data points) as a function of redshift. These data are computed as the average of the radial profiles for the components with \(\lambda_z > 0.5\) and high metallicity. The data points are the centres of the age bins, vertical errorbars show the magnitude of the variation of each profile with radius, while horizontal errorbars show the widths of the age bins. The formation redshifts are approximated from the stellar ages assuming standard \(\Lambda{\rm CDM}\) cosmology. The size of each point is proportional to the mass contained in that bin. Literature measurements are shown as box and whiskers according to the legend, and the red dashed line shows the \(18\cdot(1+z)\) evolution from \protect\cite{wisnioski15} - though we note that these measurements are of {\em star-forming,  gaseous} disks.}
    \label{img:cosmoDisp}
\end{figure}

Our model allows us to estimate the intrinsic age-velocity dispersion relation, even though NGC 3115 is unresolved. It can be seen from \cref*{img:2dDispR} that, at fixed metallicity (the enriched components in particular), there is a mild increase in velocity dispersion with age, notably at all radii. This is quantified in \cref*{img:cosmoDisp}, for the disk region of our model. To isolate this region, we make a conservative cut in circularity of \(\lambda_z > 0.5\) in order to exclude the most pressure-supported orbits, as well as select the highest metallicity bins.
Owing to how distinct our methodology is for the derivation of these data, comparison to other works is complicated. Specifically, observations of the stellar velocity dispersion across redshift are difficult to obtain. We therefore look for qualitative comparisons to other works. For instance, \cite{wisnioski15} measured, and compiled literature observations of, the {\em gas} velocity dispersion as a function of redshift. These data are taken from a range of surveys containing galaxies with \(\log_{10} (M_\star/\Msun) \in [10.1, 11.0]\). We include these data in \cref*{img:cosmoDisp}, which are specifically from HERACLES \citep{leroy09}, DYNAMO \citep{green14}, GHASP \citep{epinat10}, PHIBBS \citep{tacconi13}, MASSIV \citep{epinat12}, OSIRIS \citep{law09}, AMAZE-LSD \citep{gnerucci11}, SINS \citep{forsterschreiber09} and zC-SINF \citep{forsterschreiber14}, and KMOS\(^{3{\rm D}}\) \citep{wisnioski15}. \cite{pillepich19} investigate both the gaseous and stellar velocity dispersion as a function of redshift in the IllustrisTNG cosmological simulations for {\em star-forming} galaxies. They find the same general redshift evolution as we see here for both gas and stars in their simulations. We emphasise here, however, that the data types, galaxy types, and techniques for measuring the velocity dispersion are not directly comparable between these three works. Both \cite{wisnioski15} and \cite{pillepich19} study ensemble population properties of gas and/or stellar kinematics in star-forming galaxies, while we consider only the stellar kinematics of a single, relatively-quiescent galaxy. Nevertheless, that the shape of this redshift evolution is so consistent between these works suggests that it is robust.\par
Interestingly, we also see in \cref*{img:2dDispR} an increase in velocity dispersion towards more metal-poor components, even at fixed age and again at all radii. This implies that metallicity has a significant contribution to the increase in \(\sigma_z\), which is obscured when considering age at fixed metallicity as in \cref*{img:cosmoDisp}, or when marginalising over all metallicities as in \cref*{img:1dDispRAge}.\par
We can thus further constrain the physical mechanism driving the age-velocity dispersion relation by leveraging the formation scenario we've established in the previous sections. Following the reasoning in \cref*{ssec:formation}, the presence of a relatively cold, yet very old disk implies an upper limit on how much internal `heating' could have occurred in this galaxy, since such heating would gradually inflate the scale height of the disk's old stellar population over the galaxy's lifetime. Furthermore, any internal dynamical interactions that drive an increase in \(\sigma_z\) over time should be completely impartial to the metallicity of those stars. This is inconsistent with the rather significant change in \(\sigma_z\) with metallicity at fixed age in our model. Finally, if we interpret at face value the absolute agreement between our stellar velocity dispersion evolution and that of gas measurements in \cref*{img:cosmoDisp}, this suggests that the stars inherited their dynamical properties from the gas and there was little-to-no subsequent evolution. Once again, this is suggestive of minimal, if any, amounts of disk heating throughout the entire evolutionary history of NGC 3115. Therefore, while we can not make strong conclusions about the true physical interpretation of the age-velocity dispersion relation, nor definitively exclude {\em any} disk heating over the lifetime of the galaxy, our model indicates that the conditions at birth have the most dominant impact on a population of star's present day velocity dispersion.
\subsection{Non-Axisymmetric Structures}
The possibility of a bar-like structure and associated resonances within NGC 3115 has been discussed in the literature \citep{kormendy92, kormendy95, guerou16}. These works all conclude that there is tentative evidence for such structures, but they can neither confirm nor refute their definitive existence. \cite{guerou16} finds many coincident kinematic features that would indicate the presence of a bar. They show, however, that a simple \(N\)-body model without a bar is able to reproduce these features. We can investigate our dynamical model for evidence of kinematic signatures indicative of a bar. We find no such evidence in the orbital composition, which is dominated by short-axis tube orbits everywhere, with only a small gradual increase in the contribution from box orbits towards the centre. Our model reproduces the positive correlation between \(V\) and \(h3\) in the central off-axis region that is usually associated with a bar, despite being built in a static gravitational potential. We conclude therefore, as \cite{guerou16}, that a bar is not strictly necessary to form this feature, but whose existence can not be conclusively ruled out by our model.\par
In general, the box/long-axis-tube orbits are a unique feature of a triaxial model, and the presence of a small fraction of these orbits towards the centre of our best-fit model implies a degree of triaxiality - and may also imply a bar. We measure a very small constant oblate-triaxiality over the full model of \(T = (1-p^2) / (1-q^2) = 0.042\), with \(p=b/a\) and \(q=c/a\), for the major, intermediate, and minor axis lengths, \(a\), \(b\), and \(c\), respectively. Despite the slight increase in non-axisymmetric orbits towards the centre, there is no corresponding increase in triaxiality. However, we note here that the potential for strong triaxiality is limited in our model, since we have fixed the intrinsic shape \(u=0.999\), and are using a projected mass model which has only a single position angle. More accurate constraints on the triaxiality would thus require relaxing both of these assumptions.\par
We therefore find that while our static triaxial model is able to fully reproduce the kinematics, conclusive evidence of a bar (or otherwise) would require a more direct modelling approach including a tumbling time-variable gravitational potential, such as the {\rm NMAGIC} made-to-measure models \citep{delorenzi07, morganti12}. More importantly for the general method presented here, \cite{zhu18} showed that even for an intrinsically barred galaxy, this triaxial \shw\ code is able to accurately recover the underlying orbital distribution for the non-bar/resonance regions.

\section{Conclusions}
In this work, we have presented the application of a conceptually-new approach for the deduction of formation histories from IFU observations of external unresolved galaxies. We extracted and presented stellar kinematics and stellar-population properties across a \(\sim 240\si{\arcsecond}\) FOV of a nearby \SZ\ galaxy, NGC 3115. By fitting detailed \shw\ orbit-superposition dynamical models to the kinematics, we defined components within the galaxy that are dynamically distinct. These components were then assigned a mean stellar age and metallicity in order to reproduce the observed stellar population properties. This combination of spatial resolution, kinematics, and stellar populations allows us establish a complete history of the formation of NGC 3115:
\begin{itemize}
    \item We find that in the early gas-rich Universe, cold streams funnelled into the core of the progenitor of NGC 3115 early in its formation. These streams caused the compaction of the bulge and its mild metallicity gradient
    \item The remaining gas from this event cooled and formed stars, which began shortly after the compaction phase. Star formation continued (though declining) through to the youngest stars on the coldest orbits and with the highest metallicity
    \item Meanwhile, many low-mass satellites were being accreted, fleshing out the halo/thick-disk with low-metallicity material, gradually converting NGC 3115 into its present-day early-type lenticular morphology
\end{itemize}
More generally, we have combined the stellar dynamics and populations in a comprehensive and self-consistent manner. This has allowed us to empirically conclude that conventional galactic decomposition techniques - with few components - are unable to simultaneously fit a galaxy's shape, kinematics, and stellar populations, unless gradients are considered. We have also determined that in the case of NGC 3115, the conditions in which a population of stars forms has the dominant effect on their observed present-day kinematics. This approach amounts to a significant step towards a completely simultaneous population-dynamical model, that will drive progress in the field of galaxy formation with remarkable detail and accuracy. In future work, we will incorporate other stellar-population parameters (namely, spatially-resolved measurements of the IMF and abundances) to further improve the accuracy of the combined model presented here.\par
This method is currently being tested on mock data in order to estimate the reliability of such an approach. Leveraging the simulations will also inform this methodology on which physical properties are the most important for discriminating between the different formation paths that built up the galaxy. Finally, the successful application of this methodology to a sample of galaxies in different environments will be able to uncover the dominating mechanism(s) of formation during these galaxies' histories.

\section*{Acknowledgements}

We thank Adrien Guerou and Eric Emsellem for providing the reduced and calibrated MUSE data cube and the photometric MGE model, as well as for the stimulating discussion. We thank Emily Wisnioski for providing the literature measurements for \cref*{img:cosmoDisp}. RMcD is the recipient of an Australian Research Council Future Fellowship (project number FT150100333). LZ acknowledges support from Shanghai Astronomical Observatory, Chinese Academy of Sciences under grant NO.Y895201009. GvdV acknowledges funding from the European Research Council (ERC) under the European Union's Horizon 2020 research and innovation programme under grant agreement No 724857 (Consolidator Grant ArcheoDyn). We acknowledge support from the Australia-Germany Joint Research Cooperation Scheme funded by Universities Australia and the German Academic Exchange Service (DAAD). This work makes use of the \tfo{queenebee} compute cluster at Max-Planck-Institut f\"ur Astronomie, and the \tfo{OzStar} supercomputer at Swinbourne University. The software used for data analysis include \tso{matplotlib} \citep{matplotlib07}, \tso{AstroPy} \citep{astropy13}, and the \tso{SciPy} collection \citep{scipy01}.  We finally thank the anonymous referee, whose feedback greatly improved the clarity of this work.




\bibliographystyle{mnras}
\bibliography{3115} 


%
\appendix

\section{Mass Surface Density MGE}\label{app:MGET}
We present in \cref*{tab:MGET} the MGE fit to the mass surface-density `image' described in \cref*{sssec:stellarMassModel}, which requires \(16\) Gaussians in order to accurately describe the mass distribution.
\begin{table}
    \sisetup{group-digits=false}
    \centerline{
        \(\begin{tabu}{
            S[table-number-alignment = center]
            S[table-number-alignment = center]
            c
        }
            \mathrm{Surface\ Density}\ [\si{\Msun pc\tothe{-2}}] & \mathrm{Dispersion}\ [\si{arcsec}] & \mathrm{Axis\ Ratio}\\ \hline\hline
            5623394.70903 & 0.03032 & 0.77577\\
            357987.40788 & 0.11242 & 0.91562\\
            172654.34365 & 0.23796 &    0.85209\\
            234126.41656 & 0.48308 &    0.13169\\
            111845.72296 & 0.74309 &    0.68875\\
            140311.08130 & 1.19400 &    0.16491\\
            100124.68062 & 2.01498 &    0.51298\\
            38315.21897 &  3.84575 &    0.56651\\
            10990.86889 &  8.73513 &    0.59807\\
            12437.27038 &  14.37909 &   0.08915\\
            4097.35539 &   16.31765 &   0.48822\\
            2438.79282 &   26.88008 &   0.12718\\
            1679.17630 &   44.55499 &   0.31224\\
            465.76916 &    77.95736 &   0.38720\\
            175.55995 &    120.52598 &  0.55757\\
            21.59472 &     249.74553 &  0.71771
        \end{tabu}\)
     }
     \caption{The counts, widths, and axis ratios of the Gaussians in our \mgeT\ model.}
     \label{tab:MGET}
\end{table}

\section{Regularisation in the Dynamical Star-Formation History}\label{app:regul}
By fitting the spatially-resolved luminosity-weighted maps of stellar-population properties (described in \cref*{sssec:fsksp}) with a given number of dynamical components, there is inevitably some level of degeneracy between solutions to \cref*{eq:matrix}, which increases with increasing \(N_{\rm comp.}\). In order to reduce the impact of this degeneracy on the analyses that followed, we implemented a linear regularisation into the BVLS fit. This regularisation, as is widely used in astrophysical problems, penalises solutions which vary sharply in the parameter-space under consideration. For instance, regularisation in the context of \shw\ dynamical models would penalise solutions which have significantly different contributions from neighbouring orbits that have similar physical properties. The motivation for this is that most physical systems should undergo changes in their properties smoothly across physical parameters, rather than discretely.\par
In our application of regularisation to \cref*{eq:matrix}, we wish to impose a smoothness in the distribution of ages and metallicities that are assigned to the dynamically-selected components. In this way, we prefer solutions (which would otherwise be degenerate) that assign similar stellar-population properties to dynamical components that have similar physical properties.\par
\begin{figure}
    \centerline{
        \includegraphics[width=\columnwidth]{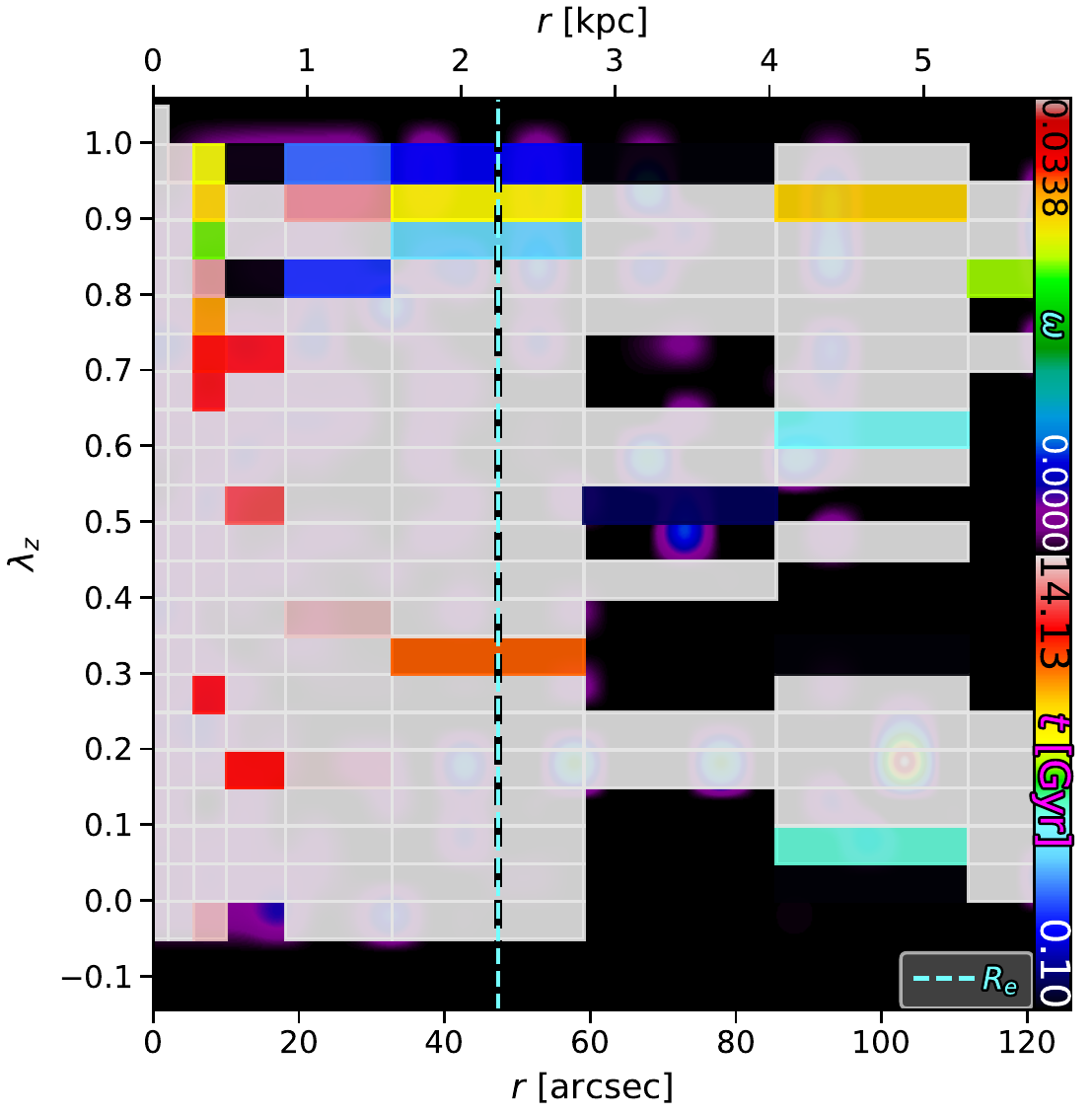}
    }
    \centerline{
        \includegraphics[width=\columnwidth]{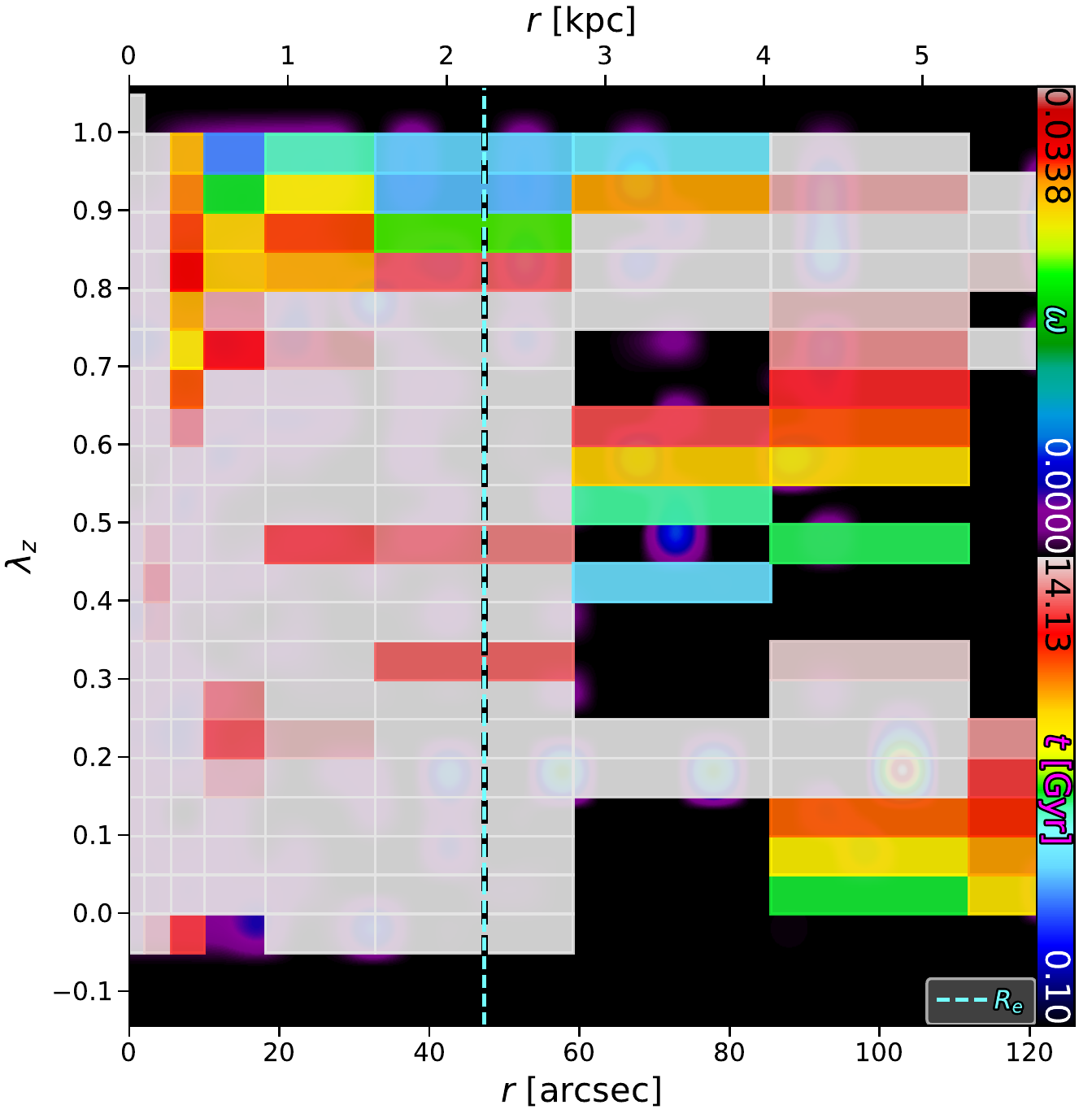}
    }
    \caption{The circularity phase-space, as shown in \protect\cref*{img:lzCut}, but where the components are additionally coloured by their assigned mean stellar age. The top panel shows an unregularised fit, while the bottom panel shows the regularisation that was used throughout this work.}
    \label{img:regulCLZ}
\end{figure}

For the regularisation, we minimise a linear approximation (as it is a linear problem) to the integral of the second-order Laplacian. Given in \S 19.5 of \cite{press07}, this can be expressed as the following:
\begin{flalign}\label{eq:Lapl}
    &&\int \left[ \vec{\phi}^{\prime\prime}(x)\right]^2 \Dif x &\propto \sum\limits_{\mu=0}^{N_{\rm comp.}-3} \left[-\phi_\mu + 2 \phi_{\mu+1} - \phi_{\mu+2}\right]^2 &&
    \intertext{for the solution vector of unknowns, \(\vec{\phi}\), of length \(N_{\rm comp.}\).}
\end{flalign}
This formalism assumes that sequential \(\mu\) are adjacent in the physical parameters of interest - in the case of our dynamically-selected components, this means neighbouring in \(\lambda_z\) and \(R\). However, as seen in \cref*{img:lzCut}, this is not the case for some components; where sequential labels wrap around to the next column, they have significantly different physical properties. In order to retain the information about which components are truly adjacent in physical properties, and avoid regularising over non-neighbouring components, we instead construct a \(2{\rm D}\) and \(3{\rm D}\) regularisation matrix for the dynamical components and orbits, respectively (which are defined by 2 and 3 parameters, respectively). This matrix preserves the second-order Laplacian given in \cref*{eq:Lapl} for each dimension, and is unique for every set of \([\mu, \mu+2]\). In order to apply the matrix to the BVLS fit, it is flattened into a single row in the same way in which the components and orbits are reduced to a single dimension along the \(N_{\rm comp.}\) axis of \cref*{eq:matrix}. In this way, the location of the regularisation constraints preserves the memory of which components are neighbouring in physical parameters. Since \cref*{eq:Lapl} affects only 3 components, this operation is repeated for each set of \([\mu, \mu+2]\), adding a row to \cref*{eq:matrix} each time. In practise it is implemented in an analogous fashion to what is described in \cite{cappellari16}. The effect of this is illustrated in \cref*{img:regulCLZ}, which compares a completely free, unregularised fit with a regularised fit which was used for the results presented in this work.
This figure illustrates that the regularisation does indeed act in the desired way, by producing `smoother' variations between the grid cells, compared to the unregularised fit. Importantly, there is a statistically-insignificant difference between the \(\chi^2\) values of the two fits, implying the regularisation is indeed acting only to break the degeneracy.

\section{Marginalised Intrinsic Velocity Dispersion Profiles}\label{app:sigma_z}
\begin{figure}
    \centerline{
        \includegraphics[width=\columnwidth]{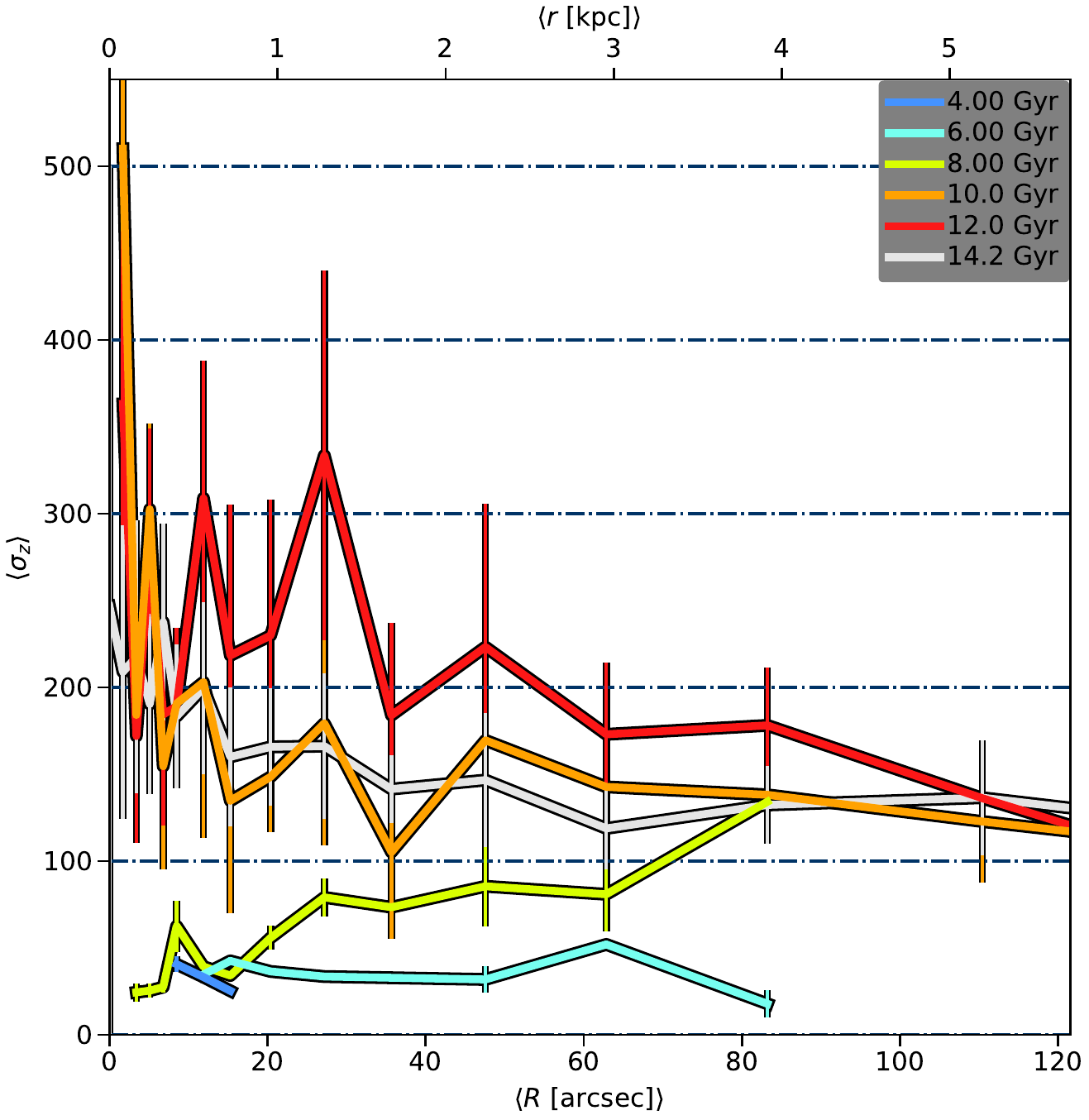}
    }
    \caption{Radial profiles of the intrinsic vertical velocity dispersion for dynamical components binned by their mean stellar age (averaged over metallicities). The absolute stellar mass of each bin is given in \cref*{img:sfhAge}. As with \protect\cref*{img:2dDispR}, the variance within the annuli are shown as errorbars.}
    \label{img:1dDispRAge}
\end{figure}

\begin{figure}
\centerline{
        \includegraphics[width=\columnwidth]{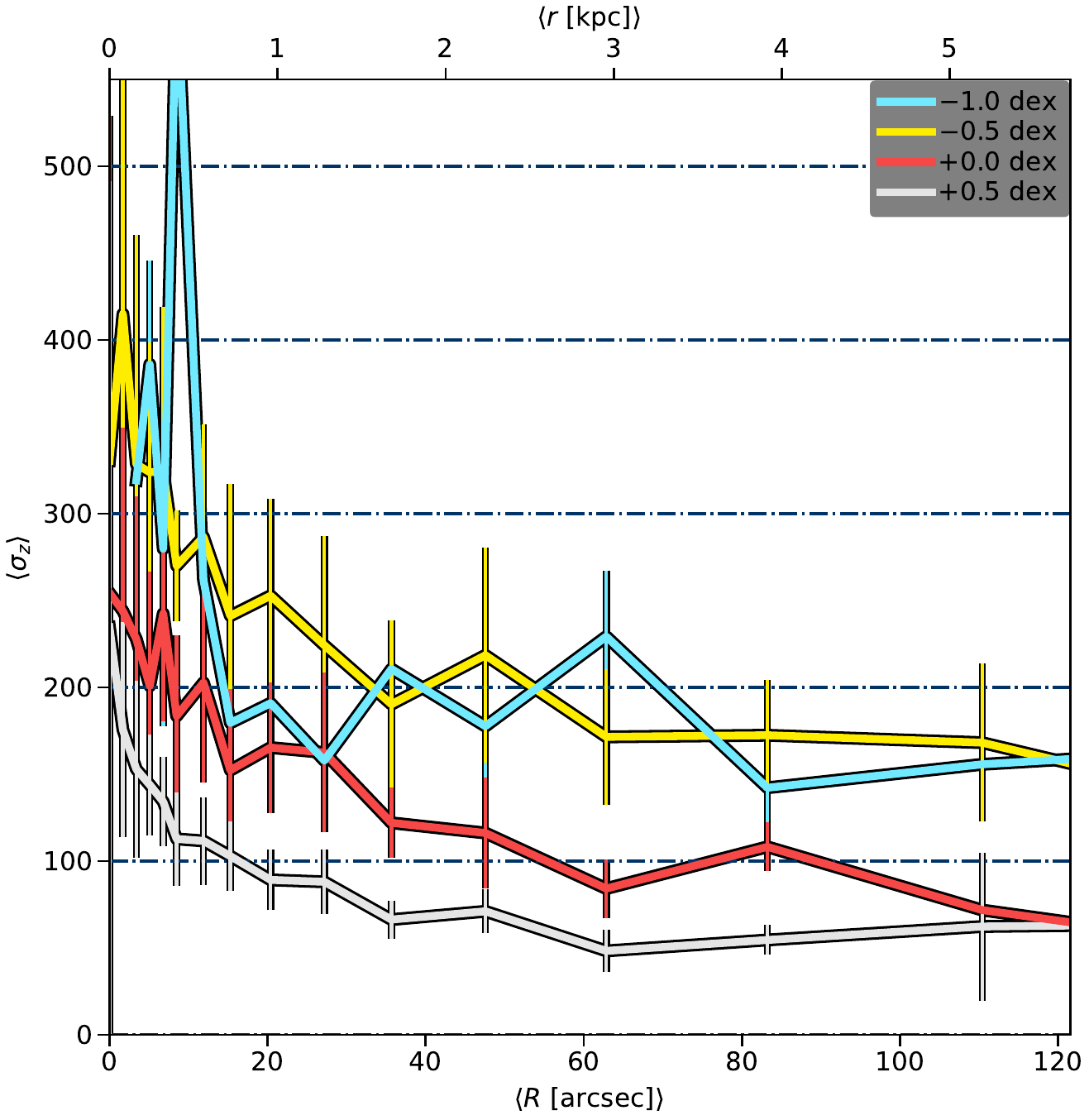}
    }
    \caption{As \protect\cref*{img:1dDispRAge}, but binned by mean stellar metallicity (averaged over ages).The absolute stellar mass of each bin is given in \cref*{img:sfhMetal}.}
    \label{img:1dDispRMetal}
\end{figure}

In \cref*{img:1dDispRAge,img:1dDispRMetal}, we present the radial profiles of the intrinsic vertical velocity dispersion for the dynamically-selected components of our model, binned only in age, and only in metallicity, respectively.

%


\bsp	
\label{lastpage}
\end{document}